\documentclass[a4paper,12pt,twoside]{article}

\usepackage[T1]{fontenc}
\usepackage[utf8]{inputenc}

\usepackage{graphicx}
\usepackage{grffile}  
\usepackage{array}
\usepackage{multirow}
\usepackage{float}

\usepackage[left=2cm,right=2cm,top=2cm,bottom=4cm]{geometry}

\usepackage[
linktocpage=true,
  colorlinks   = true,    
  urlcolor     = blue,    
  linkcolor    = blue,    
  citecolor    = red      
]{hyperref}
\usepackage{orcidlink} 

\usepackage{amsmath}

\usepackage{xspace}

\usepackage{color}

\usepackage[
backend=biber,
bibencoding=utf8,
style=numeric-comp,
sorting=none
]{biblatex}
\renewbibmacro{in:}{}

\usepackage[affil-it]{authblk}

\author[1]{Jan Kalinowski\orcidlink{0000-0001-5618-0141}}
\author[2]{Wojciech Kotlarski\orcidlink{0000-0002-1191-6343}}
\author[1]{Krzysztof Mękała\orcidlink{0000-0003-4268-508X}}
\author[1]{Paweł Sopicki\orcidlink{0000-0001-9070-1756}}
\author[1*]{Aleksander Filip Żarnecki\orcidlink{0000-0001-8975-9483}}

\title{Sensitivity  of future linear e$^+$e$^-$ colliders
       to processes of dark matter production with light mediator exchange}

\affil[1]{Faculty of Physics, University of Warsaw, Poland}
\affil[2]{Institut f\"ur Kern- und Teilchenphysik, TU Dresden, Germany}
\affil[*]{filip.zarnecki@fuw.edu.pl}

\newcommand{\ft}{\ensuremath{\text{f}^\gamma_\text{T}}\xspace}

\newcommand{\whizard}{\textsc{Whizard}\xspace}

\newcommand{\delphes}{\textsc{Delphes}\xspace}

\graphicspath{ {./plots/} }

\begin{document}

\maketitle

\vspace{2cm}

\begin{abstract}

  As any $e^+e^-$ scattering process can be accompanied by a hard
  photon emission from the initial state radiation, the analysis of
  the energy spectrum and angular distributions of those photons can
  be used to search for hard processes with an invisible final state.
  Thus high energy $e^+e^-$ colliders offer a unique possibility
  for the most general search of dark matter (DM) based on the
  mono-photon  signature.
  We consider production of DM particles at the International Linear
  Collider (ILC) and Compact Linear Collider (CLIC) experiments via a
  light mediator exchange. 
  Detector effects are taken into account within the  \delphes
  fast simulation  framework.
  Limits on the light DM production in a simplified model are set as
  a function of the mediator mass and width based on the expected
  two-dimensional distributions of the reconstructed mono-photon events.
  The experimental sensitivity is extracted in terms of the
  DM production cross section.
  Limits on the mediator couplings are then
  presented for a wide range of mediator masses and widths.
  For light mediators, for masses up to the centre-of-mass energy of the
  collider, coupling limits derived from the mono-photon analysis are
  more stringent than those expected from direct resonance searches
  in decay channels to SM particles.

\end{abstract}

\newpage


\section{Introduction}

Excess of mono-photon events is considered as one of possible
signatures of DM production at high energy $e^+e^-$ colliders.
One of the advantages of this approach is that the photon radiation from
the initial state can be described within the Standard Model and
depends only indirectly on the DM production mechanism.
Prospects for DM searches at future  $e^+e^-$ colliders were already
studied in the past,  most recent results for CLIC and ILC are presented in
\cite{deBlas:2018mhx,Blaising:2021vhh} and \cite{Habermehl:2020njb},
respectively.  
However, most of the studies focused on simplified model scenarios where the
mediator mass $M_Y$ was large, $M_Y \gg \sqrt{s}, \sqrt{|t|}$, in which case
an Effective Field Theory (EFT) description was valid.

Considered in this study is an intermediate mediator mass range,
where the mediator might still be resolved at energies accessible at
high energy ILC or CLIC. 
We use results of our recent work \cite{Kalinowski:2020lhp} for
precise modeling of mono-photon processes with \whizard
\cite{Moretti:2001zz,Kilian:2007gr}. 
We propose a novel analysis approach where the experimental sensitivity to
DM production, $e^+e^- \rightarrow \chi \, \chi$, is studied as a
function of the mediator mass and width based on the expected
two-dimensional distributions of the reconstructed mono-photon events,
$e^+e^- \rightarrow \chi \, \chi \, \gamma$. 
Limits presented in terms of the light DM production
cross section are expected to be least dependent on the model details.

Mono- and multi-photon events were studied in LEP experiments, for
collision energies up to $\sqrt{s} =
209$\,GeV~\cite{Heister:2002ut,Abdallah:2003np,Achard:2003tx,Abbiendi:2000hh}.
In particular, the observed numbers of mono-photon events were used to
set limits on the number of light neutrino flavours. 
On the other hand, assuming three active neutrino flavours, upper
limits of ${\cal O}$(100 fb) were derived on new, exotic processes with
production of an invisible final state and a single photon.
Experiments at future $e^+e^-$ colliders will allow to significantly
improve these limits and extend them to higher mediator masses.

The paper is structured as follows.
In section 2, we describe the simplified model used for modeling DM
production processes.
In section 3, the framework used for background and signal event
simulation is presented as well as the approach used to set limits on
the signal cross sections.
Constraints on DM production processes expected at 
500\,GeV ILC and 3\,TeV CLIC are discussed in section 4.
Section 5 presents our conclusions. 


\section{Computational framework}

\subsection{Simplified DM model}

We consider a simplified DM model which covers most popular scenarios of
dark matter pair-production at $e^+e^-$ colliders.
In this model the dark matter particles, $\chi_i$, couple to the
SM particles via the mediator, $Y_j$.
Each simplified scenario is characterized by one mediator 
and one dark matter candidate 
from the set listed in Tab.~\ref{tab:model_content}.
 \begin{table}[tb]
 \centering
 \begin{tabular}{|c|c|c|c|c|c|}
 \hline
 & \multicolumn{1}{c|}{particle} & \multicolumn{1}{c|}{mass} & \multicolumn{1}{c|}{spin} & \multicolumn{1}{c|}{charge} & \multicolumn{1}{c|}{self-conjugate}\\
 \hline
 \hline
 \parbox[t]{2mm}{\multirow{3}{*}{\rotatebox[origin=c]{90}{mediator\,}}}
 & $Y_R$ & $M_{Y_R}$ & 0 & 0 & yes\\[1mm]
 & $Y_V$ & $M_{Y_C}$ & 1 & 0 & yes\\[1mm]
 & $T_C$ & $M_{T_C}$ & 0 & 1 & no\\
 \hline
 \hline
 \parbox[t]{2mm}{\multirow{5}{*}{\rotatebox[origin=c]{90}{DM}}}
 & $\chi_R$ & $m_{\chi_R}$ & 0 & 0 & yes\\
 & $\chi_C$ & $m_{\chi_C}$ & 0 & 0 & no\\
 & $\chi_M$ & $m_{\chi_M}$ & $\tfrac{1}{2}$ & 0 & yes\\
 & $\chi_D$ & $m_{\chi_D}$ & $\tfrac{1}{2}$ & 0 & no\\
 & $\chi_V$ & $m_{\chi_V}$ & 1 & 0 & yes\\
 \hline
 \end{tabular}
 \caption{Summary of mediators and DM candidates included in the
   simplified DM model considered.  \label{tab:model_content}}
 \end{table}
The interaction between DM and the electrons can be mediated by a real
scalar $Y_R$ or a real vector $Y_V$, with the Lagrangian describing mediator
coupling to electrons given by
\begin{eqnarray*}
  \mathcal{L}_{eeY} & \ni & 
     \bar{e} (g_{e Y_R}^1 + \imath \gamma^5 g_{e Y_R}^5) e Y_R +
   \bar{e} \gamma_\mu (g_{e Y_V}^1 + \gamma^5 g_{e Y_V}^5) e Y_V^\mu
   \, .
  \end{eqnarray*}
Only real values of couplings $g_{e Y_R}^1$, $g_{e Y_R}^5$, $g_{e Y_V}^1$ and
$g_{e Y_V}^5$ are taken into account, and, depending on the coupling choice,
six different scenarios are considered in the presented study, 
as listed in Tab.~\ref{tab:model_scenarios}.
 \begin{table}[tb]
 \centering
 \begin{tabular}{|l|c|c|c|c|c|}
 \hline
scenario & mediator & $g_{e Y_R}^1$ & $g_{e Y_R}^5$ & $g_{e Y_V}^1$ &
$g_{e Y_V}^5$ \\
 \hline
 \hline
Scalar & $Y_R$ & $g_{eeY}$ & 0 & 0 & 0 \\
Pseudo-scalar & $Y_R$ & 0 &  $g_{eeY}$ & 0 & 0 \\ \hline\hline
Vector & $Y_V$ & 0 & 0 & $g_{eeY}$ & 0  \\
Pseudo-vector & $Y_V$ & 0 & 0 & 0 &  $g_{eeY}$  \\
V$-$A coupling & $Y_V$ & 0 & 0 & $g_{eeY}$ & $- g_{eeY}~~$  \\
V+A coupling & $Y_V$ & 0 & 0 & $g_{eeY}$ &  $g_{eeY}$  \\
 \hline
 \end{tabular}
 \caption{Summary of mediator scenarios considered in the presented
   study.  \label{tab:model_scenarios}} 
 \end{table}
In addition, five DM candidates can be considered: a real scalar
$\chi_R$, a complex scalar $\chi_C$, a Majorana fermion $\chi_M$ and a Dirac
fermion $\chi_D$, and a real vector $\chi_V$.
  We choose to investigate the sensitivity of future e$^+$e$^-$
  colliders in terms of the DM production cross section.
  For DM particle mass significantly below half of
  the mediator mass, pair-production cross section can be given in
  terms of the mediator partial widths (to electrons and DM particle),
  its total width and mass. 
  Since we assume that the total width is dominated by the DM partial width,
  cross section dependence on the DM particle couplings is
  absorbed in the total mediator width and the limits extracted
  for fixed mediator mass and width hardly depend on the DM
  particle type or coupling structure. 
Therefore, only the Dirac fermion, $\chi_D$, is considered as the DM particle,
and its interactions with possible mediators are described by
\begin{eqnarray*}
  \mathcal{L}_{\chi\chi Y} & \ni &
  \bar{\chi}_D (g_{\chi_D Y_R}^1 + \imath \gamma^5 g_{\chi_D Y_R}^5) \chi_D Y_R +
   \bar{\chi}_D \gamma_\mu (g_{\chi_D Y_V}^1 + \gamma^5 g_{\chi_D Y_V}) \chi_D Y_V^\mu
\, ,
\end{eqnarray*}
where, for simplicity, for each scenario described in
Tab.~\ref{tab:model_scenarios} the structure of mediator couplings to DM
fermions is assumed to be the same as to SM fermions.
The DM particle type and its coupling structure becomes
relevant only when the cross section limits are translated to the limits
on the product of mediator couplings.\footnote{For example,
limits on the product of mediator couplings,
$g_{eeY} g_{\chi\chi Y}$, presented in Sec.~\ref{sec:results} 
decrease by a factor of $\sqrt{2}$, if Majorana fermion DM is assumed.}
Moreover, to simplify the discussion in terms of the mediator properties, 
we fix the mass of the DM fermion to $m_\chi = 50$\,GeV throughout the
paper.\footnote{For light mediator scenarios,
  DM particle mass has marginal impact on the analysis results as long
  as it is significantly below half of the mediator mass.}
On the other hand, if an excess of mono-photon events is
observed, determination of the DM type and coupling structure
affecting the observed photon distribution is possible
\cite{Choi:2015zka}, which however goes beyond the scope of the
current study.

The model defined by the Lagrangian described above, also including
other DM particle types shown in Tab.~\ref{tab:model_content}, has
been encoded into
\textsc{FeynRules}~\cite{Christensen:2008py,Alloul:2013bka} and
exported in \textsc{UFO} format~\cite{Degrande:2011ua}, which was then
used as an input to \whizard
\cite{Moretti:2001zz,Kilian:2007gr}.
More details as well as the \textsc{UFO} file of the model can be found in
\cite{SimpDMwiki}.

\subsection{Theoretical and experimental constraints}

This study focuses on scenarios where the mediator might
be produced on-shell at energies accessible at high energy ILC or
CLIC.
For the validity of the cross section calculations and event sample
generation in \whizard (see sec.~\ref{sec:whizard}), we have to assume
that the total width of the mediator is smaller than its mass (so the
mediator itself is well defined as a particle) and its couplings
are small (so perturbative approach is applicable).
The first condition is fulfilled by fixing the ratio of mediator width to
mediator mass in the considered analysis scenarios and restricting the
study to the range $\Gamma/M \le 0.5$.
To ensure validity of perturbative calculations and to respect the
experimental constraints described below, coming from LEP and LHC
experiments, we assume mediator coupling to electrons of $g_{eeY} =
0.01$ for reference cross section calculations and
signal sample generation.  

Radiative DM pair-production scenario described above was not
addressed in the analysis of mono-photon events at LEP
\cite{Heister:2002ut,Abdallah:2003np,Achard:2003tx,Abbiendi:2000hh}. 
However, limits on the mono-photon production cross section at
$\sqrt{s} =$ 205\,GeV presented in \cite{Abdallah:2003np} can be
compared with generator-level predictions of our model.
Limits of 50$-$150\,fb   
on the cross section for single photon production in the
central detector region, $45^\circ < \theta_\gamma < 135^\circ$,
correspond to the limits on the mediator coupling to electrons of the
order of 0.01$-$0.02 extending almost up to the kinematic limit,
$M_Y \le$ 200\,GeV.

The strategy employed by ATLAS and CMS is to show exclusion
limits in the DM mass versus mediator mass plane for selected
benchmark scenarios proposed in \cite{Albert:2017onk}.
For light DM states, results from the ATLAS collaboration
\cite{ATLAS:2020uiq}, looking for the dark matter production in
association with an energetic photon at $\sqrt{s} =$ 13\,TeV,
exclude mediator masses up to 920$-$1470\,GeV, depending on the
scenario.
However, scenarios \cite{Albert:2017onk} assume relatively strong
mediator coupling to quarks, $g_{qqY}=$ 0.1 or 0.25.
For quark coupling values below 0.03, no constraints on the light mediator
scenarios can be set in any of the considered analysis approaches
\cite{ATLAS:2019wdu}. 
Also CMS results \cite{CMS:2021ctt} on the resonance search in
high-mass dilepton final states at $\sqrt{s} =$ 13\,TeV indicate that
mediator masses up to about 3\,TeV can be excluded for large mediator
coupling values to quarks and leptons, $g_{qqY}=g_{llY}=0.1$.
However, for weaker lepton coupling, $g_{llY}=0.01$,
scenarios with light DM particle and mediator mass below 1\,TeV are
still not excluded.

\subsection{Search strategy}
\label{sec:strategy}

We propose a novel approach to the DM searches at
colliders, where the experimental sensitivity is defined in terms of
both the mediator mass and mediator width.
Limits on the light DM production cross section can be set
based on the 
two-dimensional distributions of the reconstructed mono-photon events.
This approach is more model independent than the approaches presented
so far, in which given mediator coupling values to SM and DM
particles were assumed.
Also, as the observed photons predominantly come from initial state
radiation, extracted cross section limits hardly depend on
the details of the BSM scenario.

To simplify the presentation, we consider models with fixed ratio of
mediator width to mediator mass.
We consider ``narrow'' mediator scenarios with $\Gamma/M=$ 0.01 and
0.03, ``medium'' with  $\Gamma/M=$ 0.1  and ``wide'' with $\Gamma/M=$ 0.5. 
We assume that the mediator coupling to SM particles is small (the
reference coupling value is 0.01) so that mediator decays to SM particles
can be neglected.
Our results (cross section limits for invisible decays) remain valid
also for the case when decays to SM particles would be visible, 
but the analysis based on combination of visible and invisible decay
channels is beyond the scope of this study.

\subsection{Linear collider running scenarios}
\label{sec:running}

The baseline running scenario for the ILC assumes the integrated
luminosities of about 2\,ab$^{-1}$ at 250\,GeV and 
4\,ab$^{-1}$ at 500\,GeV, with an additional 200\,fb$^{-1}$ collected at
the top-quark pair-production threshold around 350\,GeV
\cite{Bambade:2019fyw}.
The accelerator design allows for polarisation of both $e^-$ and $e^+$ beams,
of 80\% and 30\%, respectively.
The H-20 running scenario for ILC was proposed in 
\cite{Barklow:2015tja}, to optimise the physics performance of the experiment.
It assumes
that 80\% of the total integrated luminosity at 500\,GeV will be
collected with opposite beam polarisations (1600\,fb$^{-1}$ for
left-handed electrons and right-handed positrons, $e^-_L e^+_R$, and 1600\,fb$^{-1}$ for $e^-_R e^+_L$ running, denoted
as LR and RL configurations in the following) and 20\%
with same polarisations of both beams (400\,fb$^{-1}$ for
$e^-_L e^+_L$ and 400\,fb$^{-1}$ for $e^-_R e^+_R$ running, referred
to as LL and RR runs).

The implementation plan for CLIC \cite{Aicheler:2019dhf} assumes
collecting 1~ab$^{-1}$ at the first construction stage with an energy
of  380 GeV,
while 2.5\,ab$^{-1}$ and  5\,ab$^{-1}$ are expected to be collected 
for the second and third construction stages, at 1.5\,TeV and 3\,TeV.
Only the electron beam polarisation of 80\% is included in the CLIC
baseline design.
It is assumed that 80\% of the integrated luminosity at the 3\,TeV
stage (4\,ab$^{-1}$) will be collected with negative (left-handed) electron 
beam polarisation and 20\% (1\,ab$^{-1}$) with positive polarisation
(right-handed electrons). 

Sensitivity of linear $e^+ e^-$ collider experiments to DM
pair-production is expected to increase with collision energy, both
because of the increased mass range for the on-shell mediator
production and of the assumed increase in the integrated luminosity.
Therefore, considered in the presented study are ILC and CLIC running
at the energy stages of 500\,GeV and  3\,TeV, respectively.

\subsection{Model predictions}

\begin{figure}[tb]
 \centerline{\includegraphics[width=0.49\textwidth]{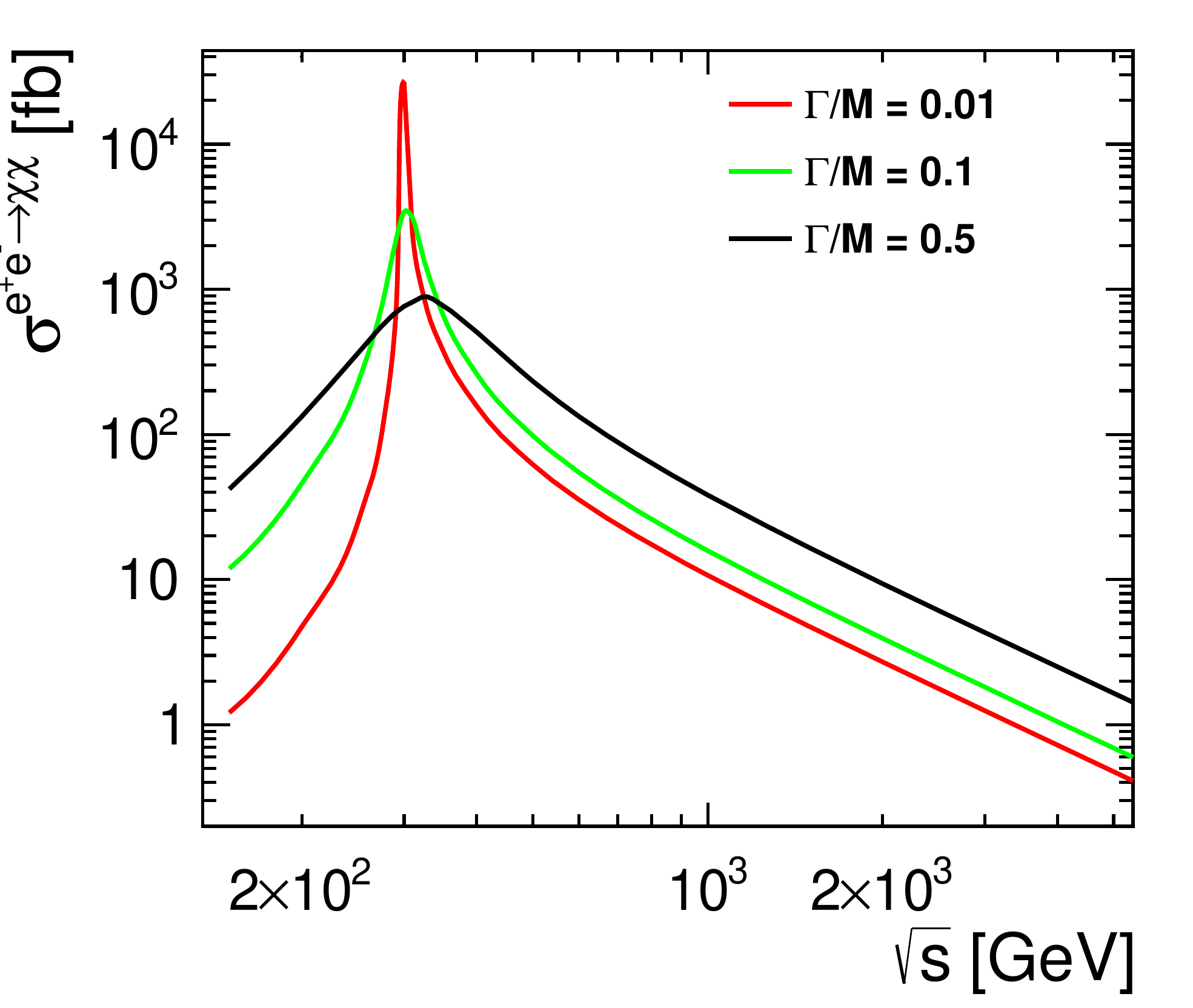}}
  \caption{Cross section for DM production in $e^+e^-$ collisions
    as a function of the collision energy $\sqrt{s}$. Expectations for
    pair production of light Dirac DM ($m_\chi$ =  50\,GeV) are shown
    for the scenario with vector mediator of 300\,GeV and different
    mediator widths, as  indicated in the plot. 
    The mediator coupling to electrons is set to 0.01.} 
  \label{fig:exp_ene}
\end{figure}

Shown in Fig.~\ref{fig:exp_ene} is the cross section for DM
production,
  $e^+e^- \rightarrow \chi\, \chi $,
as a function of the $e^+e^-$ collision energy.
Pair-production of light Dirac DM, $m_\chi$ =  50\,GeV, is considered
for scenario with the vector mediator of 300\,GeV and different
mediator widths, as described above.
Possible beam polarisation and beam energy spectra are not taken into account.
Large enhancement of DM production is possible for ``narrow'' mediator
scenario, when the beam energy is tuned to the mediator mass.
For ``light'' mediator scenarios, when the collision energy is much
higher than the mediator mass, $M_Y \ll \sqrt{s}$, production cross section decreases as $1/s$.

\begin{figure}[tb]
\includegraphics[width=0.49\textwidth]{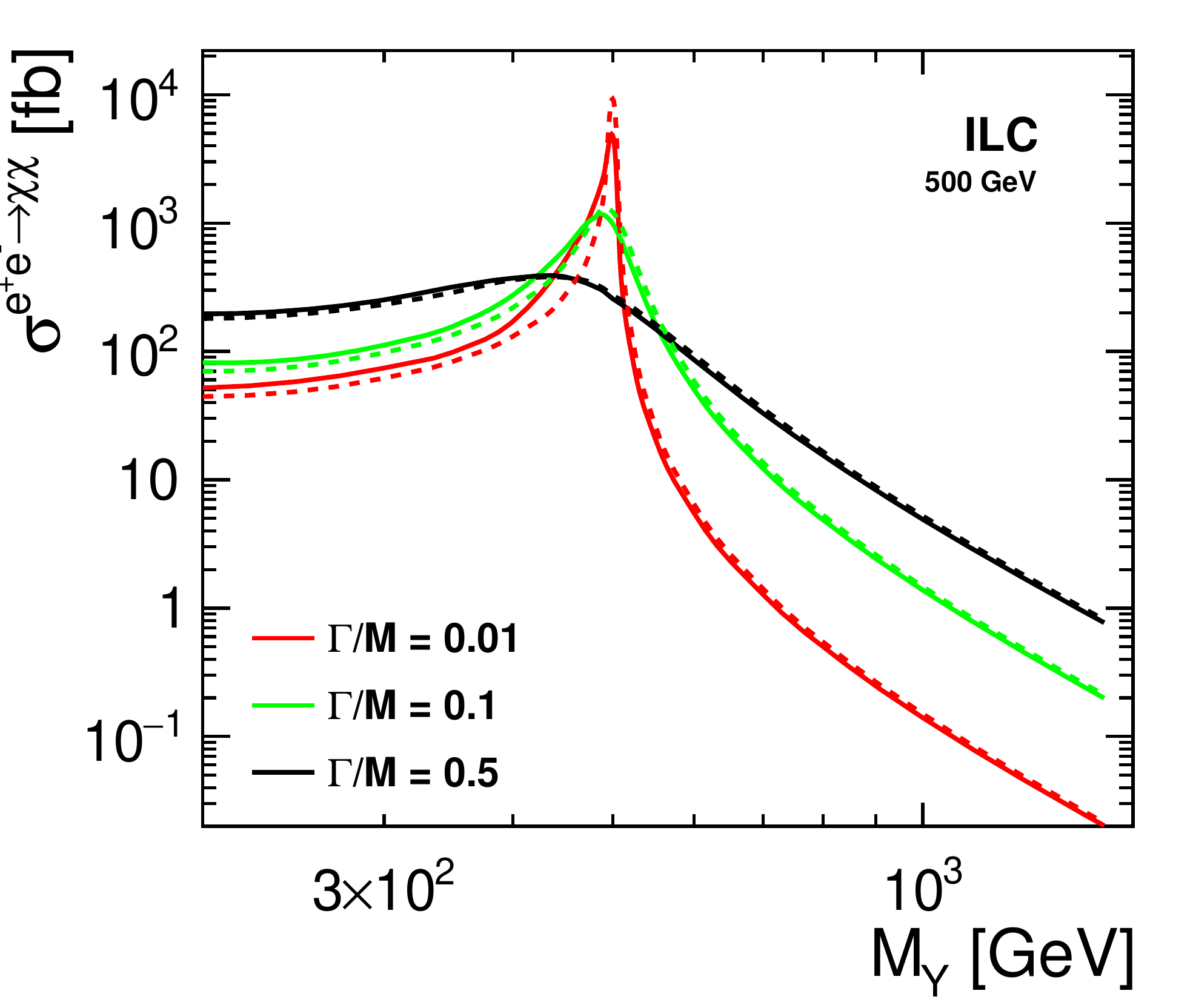}
\includegraphics[width=0.49\textwidth]{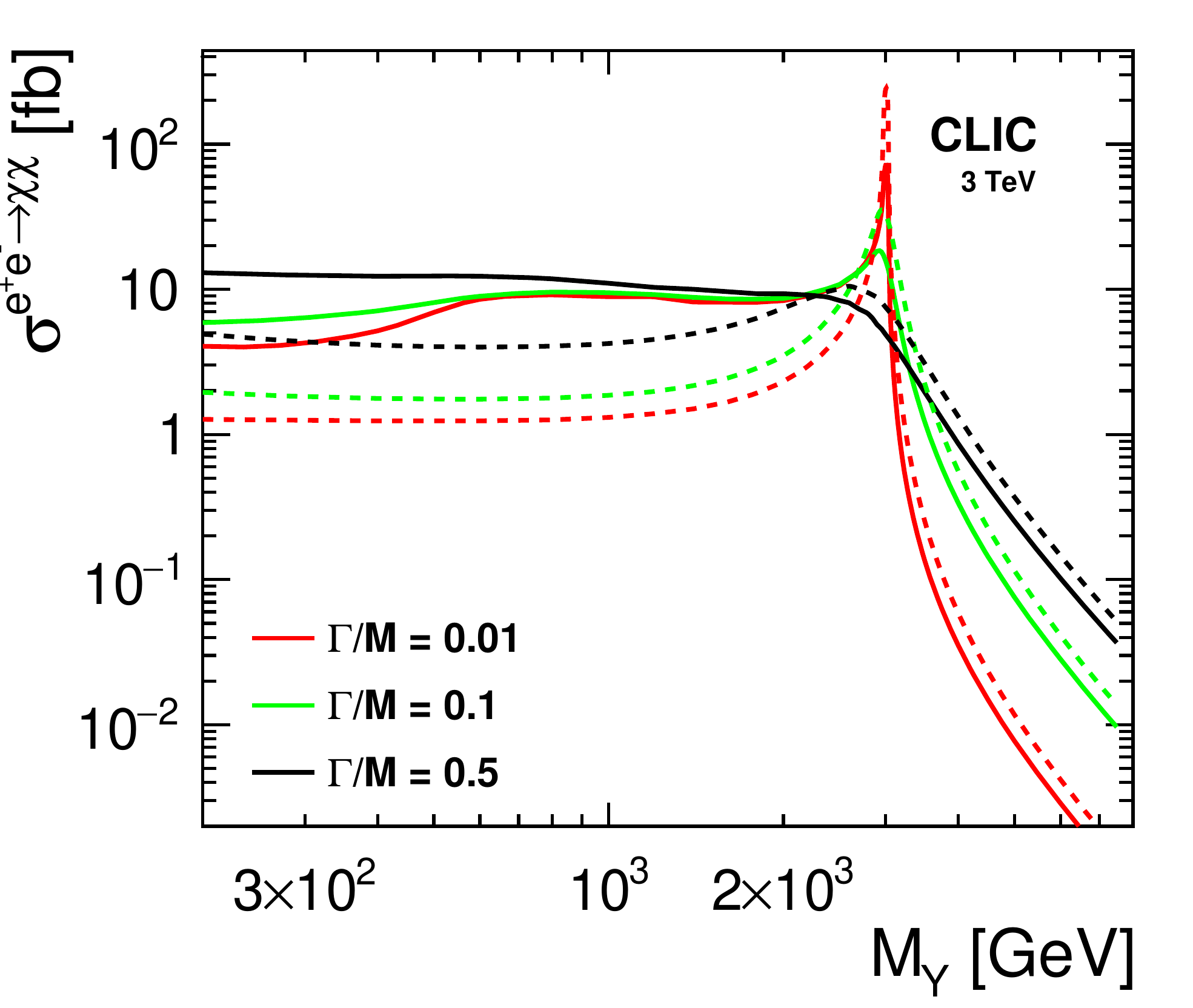}
  \caption{Cross sections for DM production at 500\,GeV ILC (left) and
    3\,TeV CLIC (right) as a function of the assumed mediator
    mass, $M_Y$. Expectations for pair production of light Dirac DM
    ($m_\chi$  =  50\,GeV) are shown for scenario with vector mediator
    and different fractional mediator widths, as indicated in the plot.
    Results obtained with the expected collider luminosity spectra
    taken into account (solid lines) are compared with results
    assuming monochromatic beam (dashed lines).
    Mediator coupling to electrons is set to 0.01.} 
  \label{fig:exp_mass}
\end{figure}

In the following, we consider scenarios with light mediator and those
for which the mediator mass and collision energy are comparable.
Shown in Fig.~\ref{fig:exp_mass} are the expected cross sections for
DM production at  500\,GeV ILC and 3\,TeV CLIC, as a function of the
assumed mediator  mass for different fractional mediator
widths.
It is interesting to note that for scenarios with light mediator
the production cross section depends very weakly on the
mediator mass.
Also indicated in Fig.~\ref{fig:exp_mass} is the impact of the
collider luminosity spectra on the expected DM production cross section.
At the ILC the effect of the luminosity spectra is significant
only for the ``narrow'' mediator scenario and in the vicinity of the
resonance.
It is much more important at CLIC where, due to the large contribution
of the low energy tail in the beam energy spectra, the cross section
for light DM production with exchange of light mediator is
significantly enhanced, up to about a factor of 5. 

\section{Data Simulation and Analysis}

\subsection{Whizard simulation}
\label{sec:whizard}

\begin{figure}[tbp]
  \includegraphics[width=0.49\textwidth]{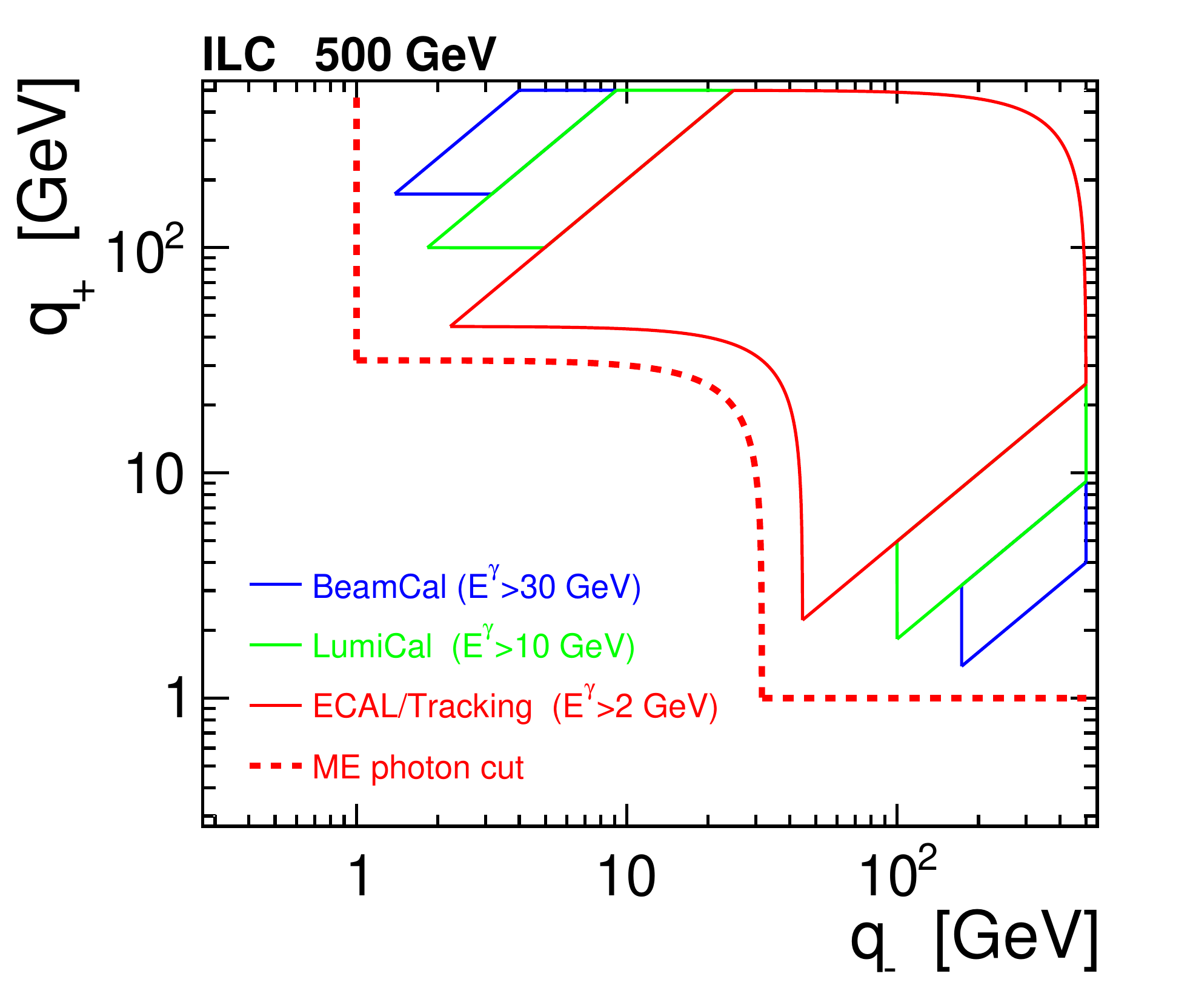}
  \includegraphics[width=0.49\textwidth]{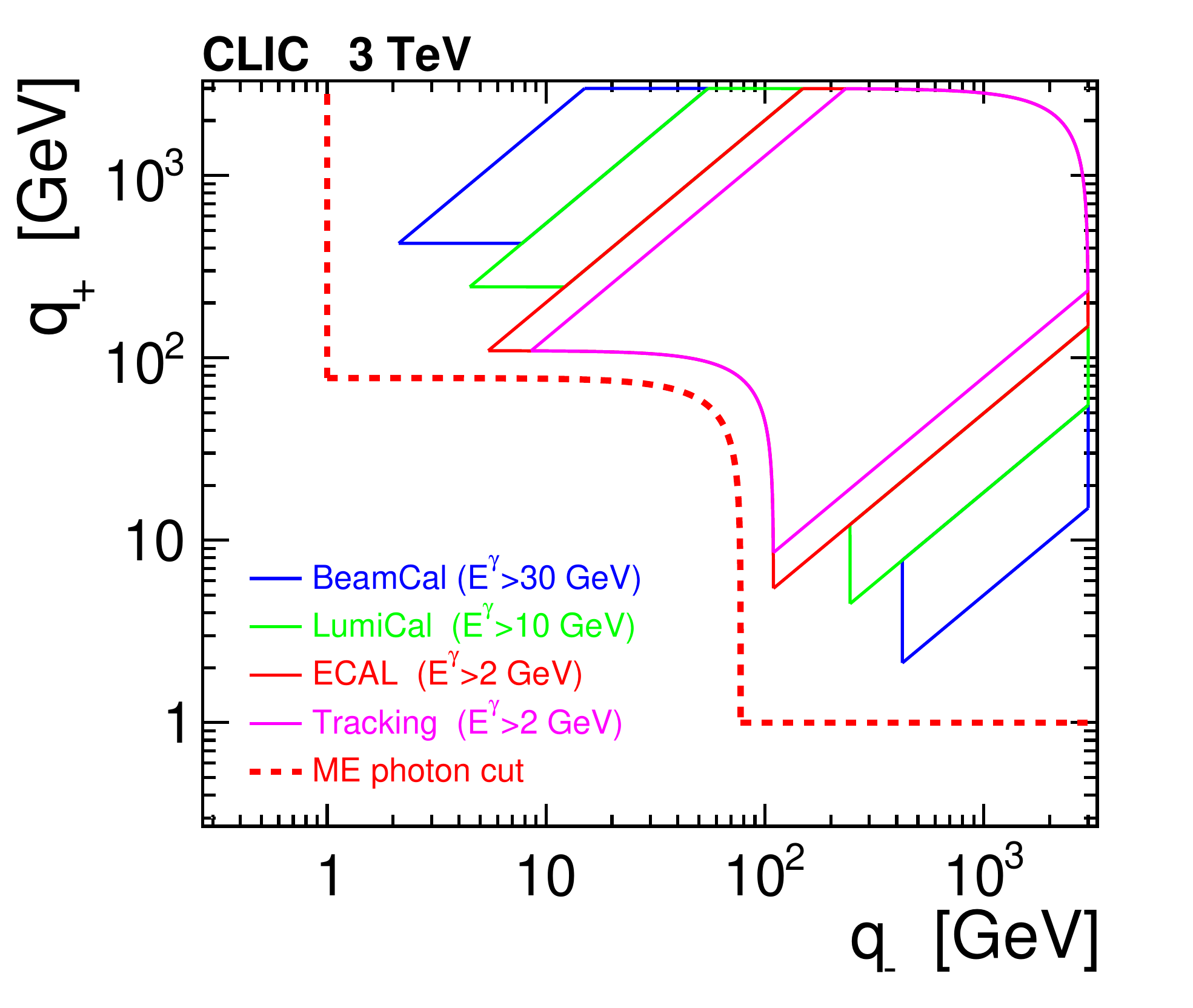}
  \caption{Detector acceptance in the $(q_+,q_-)$ plane expected for
    the future experiments at 500\,GeV ILC (left) and 3\,TeV CLIC
    (right). Red dashed lines indicate the cut
    used to restrict the phase space for ME photon generation.
   } 
  \label{fig:q_plot}
\end{figure}

Main SM background contributions to searches based on mono-photon
signature are expected to come from the radiative neutrino pair production,
$e^+e^- \rightarrow \nu \, \nu \, \gamma$, 
and the radiative Bhabha scattering,
$e^+e^- \rightarrow e^+ e^- \, \gamma$.
For precise kinematic description of photons entering the
detector, hard photon emission should be included directly in the
matrix element (ME) calculation.
On the other hand, very soft and collinear photons should still be
simulated with the parametric approach, taking into account proper
resummation of higher order corrections.
The procedure proposed in \cite{Kalinowski:2020lhp}
allows for consistent, reliable simulation of mono-photon events
in \whizard\cite{Moretti:2001zz,Kilian:2007gr} for both BSM signal
and SM background processes, based on  merging the ME calculations
with the lepton ISR structure function. 
Two variables, calculated separately for each emitted photon, are used
to describe kinematics of the emission \cite{Kalinowski:2020lhp}:
\begin{eqnarray*}
  q_{-} & = & \sqrt{4 E_{_0} E_\gamma} \cdot
  \sin{\frac{\theta_\gamma}{2}} \; , \\
  q_{+} & = & \sqrt{4 E_{_0} E_\gamma} \cdot
  \cos{\frac{\theta_\gamma}{2}} \; ,
  \end{eqnarray*}
where $E_{_0}$ is the nominal electron or positron beam energy, while $E_\gamma$
and $\theta_\gamma$ are the energy and scattering angle of the emitted
photon in question.
For the single photon emission they would correspond to the virtuality
of the electron or positron after (real) photon emission.
They are introduced to define the phase-space region where the radiated
photon can be reconstructed in the detector, as shown in Fig.~\ref{fig:q_plot}.
Detector acceptance indicated in the plot corresponds to the
considered detector models, see Sec.~\ref{sec:detsim}.
Only photons with large values of virtualities $q_-$ and $q_+$ can be
measured in the detector.
We therefore require photons generated at the ME level to have energy
above $E_{min}=1$\,GeV and $q_\pm$ virtualities above the merging scale
$q_{min}=1$\,GeV.
At the same time, to avoid double counting,
we reject the events with
any of the ISR photons passing the ME photon selection
cuts.

Only radiative events were simulated for SM background processes.
For 500\,GeV ILC (3\,TeV CLIC) we require at least one ME
photon to  be emitted above 5$^{\circ}$ (7$^{\circ}$)
from the beam axis and with a transverse momentum higher than 2\,GeV
(5\,GeV).
After the hard photon selection described above, contribution to the
radiative cross section from processes with three ME photons is at
per-cent level \cite{Kalinowski:2020lhp}.
Processes with up to three ME photons have been thus included to
match the expected precision of the measurement. 
Contribution from higher order processes is at the level of $10^{-5}$
and has been neglected.

\begin{figure}[tbp]
 \includegraphics[width=0.49\textwidth]{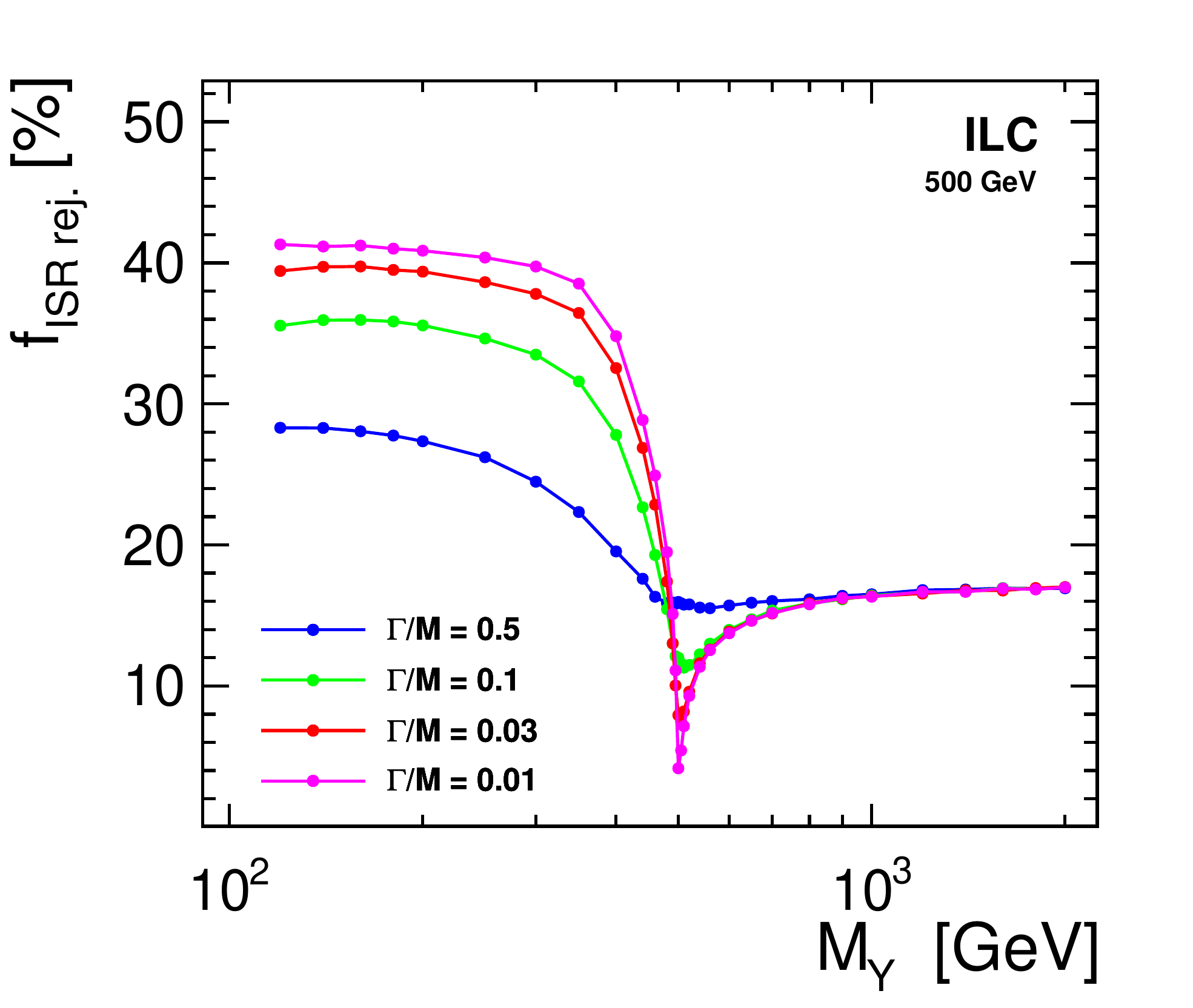}
 \includegraphics[width=0.49\textwidth]{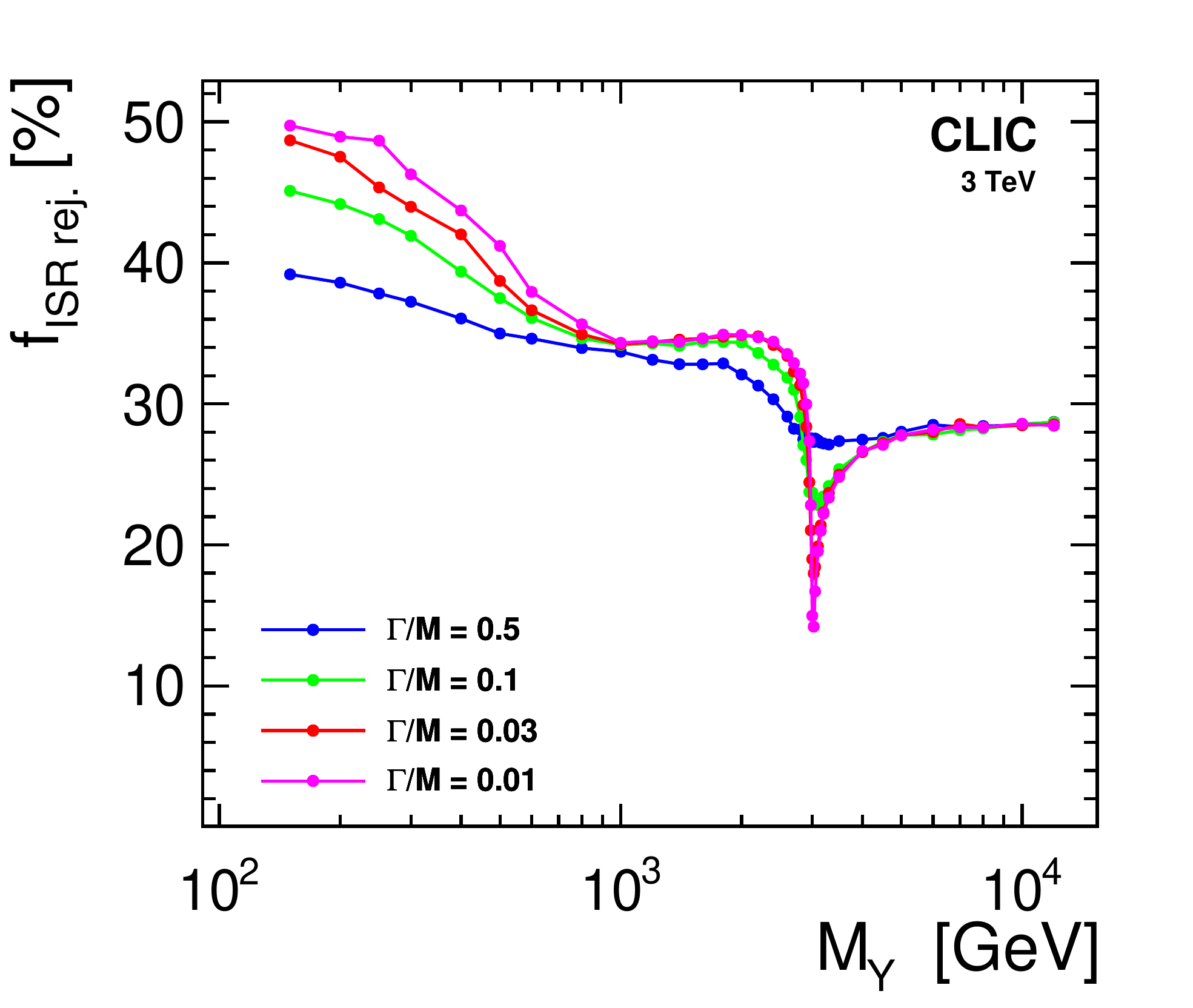}
  \caption{Fraction of \whizard events, which are removed by the ISR
    rejection procedure, as described in
    \cite{Kalinowski:2020lhp}. See text for details.
  } 
  \label{fig:isr_rej}
\end{figure}

For the considered DM pair-production scenarios,
generated signal samples included both non-radiative and radiative
processes (up to three ME photons) to ensure proper evaluation of
the total production cross section. 
The ISR rejection procedure has much stronger impact on the signal
generation than for the SM background, resulting in rejection of up to
50\% of generated events, as shown in Fig.~\ref{fig:isr_rej}.
Highest fraction of events is rejected for low mediator masses and
small mediator widths, when radiative return to the mediator mass results
in a significant enhancement of the radiative cross section.
On the other hand, for the resonant mediator production at nominal
collision energy, $M_Y \approx \sqrt{s}$, photon radiation is
significantly suppressed, and so the number of rejected events is much
smaller. 

\subsection{Detector simulation}
\label{sec:detsim}

Results presented in our previous work \cite{Kalinowski:2020lhp}
were based on simple acceptance cuts applied at the generator level.
Only the expected geometrical acceptance of the detector was
taken into account, while the reconstruction
efficiency and the detector resolution was not considered.
Proper description of the reconstruction efficiency for 
electrons scattered at low angles is crucial for the proper
modeling of the background coming from the radiative Bhabha scattering
\cite{Habermehl:2020njb}.
We therefore use the fast simulation framework
\delphes\cite{delph} in which the two detector models were implemented
recently: 
\texttt{ILCgen} for ILC running at 500\,GeV and
\texttt{CLICdet\_Stage3\_fcal},
the extended version of \texttt{CLICdet\_Stage3} \cite{Leogrande:2019qbe},
for 3\,TeV CLIC.
Both models include description of the calorimeter systems in the very
forward region: the luminosity calorimeter (LumiCal) and the beam
calorimeter (BeamCal).
Their angular acceptance extends down to 6\,mrad for
ILC and 10\,mrad for CLIC.
Implemented in the \delphes detector models are the reconstruction
efficiencies and detector resolutions expected from the
full simulation results \cite{Arominski:2018uuz,ILD:2020qve,Habermehl:2018yul}.

\subsection{Event selection}
\label{sec:selection}

For processes of DM pair-production, we expect radiative photons to be
the only final state particles in the detector.
The same signature (but with different kinematic distribution)
is expected for the SM background coming from the radiative neutrino
pair-production.
However, the situation is very different for the radiative Bhabha
scattering.
Transverse momentum of the radiated photon has to be balanced by a
scattered electron or another photon.

\begin{figure}[tbp]
  \includegraphics[width=0.49\textwidth]{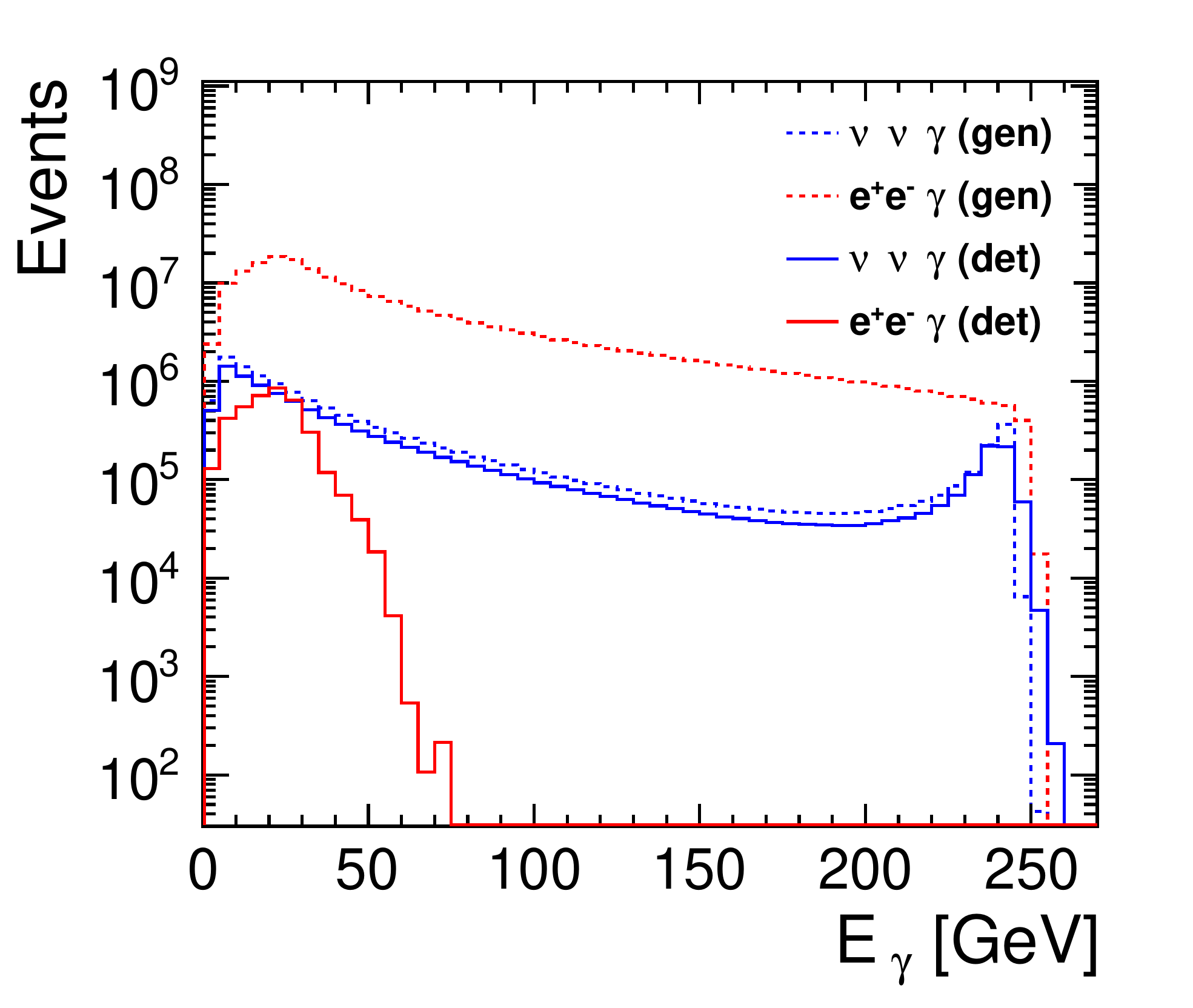}
  \includegraphics[width=0.49\textwidth]{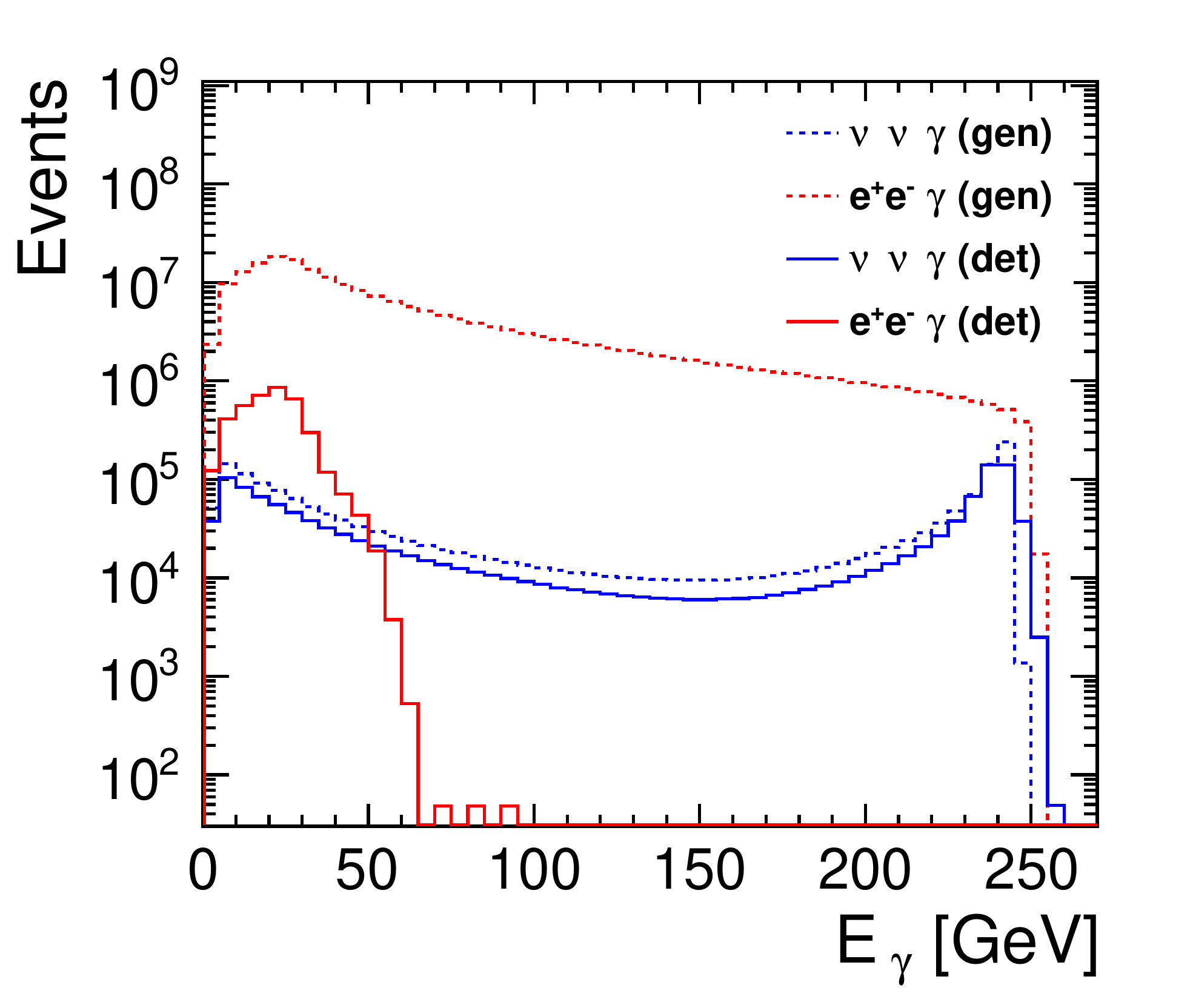}
  \caption{Energy distributions for central photons
    emitted in radiative neutrino pair production events and radiative
    Bhabha events.
    Distributions expected for selected mono-photon events, after
    detector response simulation with \delphes (solid lines), are
    compared with generator level results from \whizard (dashed
    lines).
    Numbers of events presented correspond to the integrated
    luminosity of 1.6\,ab$^{-1}$, for ILC running at 500\,GeV with
    electron/positron beam polarisation of $-80\%$/$+30\%$ (left) and
     $+80\%$/$-30\%$ (right).
  } 
  \label{fig:egamma_dist}
\end{figure}

Photons can be reconstructed with high efficiency and high purity only
in the central part of the detector, as the tracking detectors are
required to distinguish between photons and electrons (or
positrons). 
To suppress the Bhabha contribution we assume that we should observe only
a single photon in the central detector region and no other
deposits or tracks.
The mono-photon acceptance region in photon rapidity and transverse
momentum plane is defined as $|\eta_\gamma| < 2.8$, $p^\gamma_T >$3\,GeV
for 500\,GeV ILC and $|\eta_\gamma| < 2.6$, $p^\gamma_T >$10\,GeV  for
3\,TeV CLIC.
Shown in Fig.~\ref{fig:egamma_dist} are the expected photon energy
distributions for selected mono-photon events, for ILC 
running with LR and RL beam polarisations.
Generator level distributions (without mono-photon selection) are
shown for comparison.
While the selection cuts hardly affect the spectra for radiative
neutrino pair-production events,\footnote{For highest
photon energy values, $E_\gamma \approx 250$\,GeV, numbers of events after
selection cuts are higher than numbers of generated events in given
energy bin due to detector resolution effects.} 
Bhabha background is significantly suppressed.
After the selection cuts, which include veto based on BeamCal and
LumiCal response, Bhabha scattering background is limited to low
photon transverse momentum values.
SM background to mono-photon events with high photon transverse
momenta is dominated by radiative neutrino pair-production.

\begin{figure}[tbp]
 \includegraphics[width=0.49\textwidth]{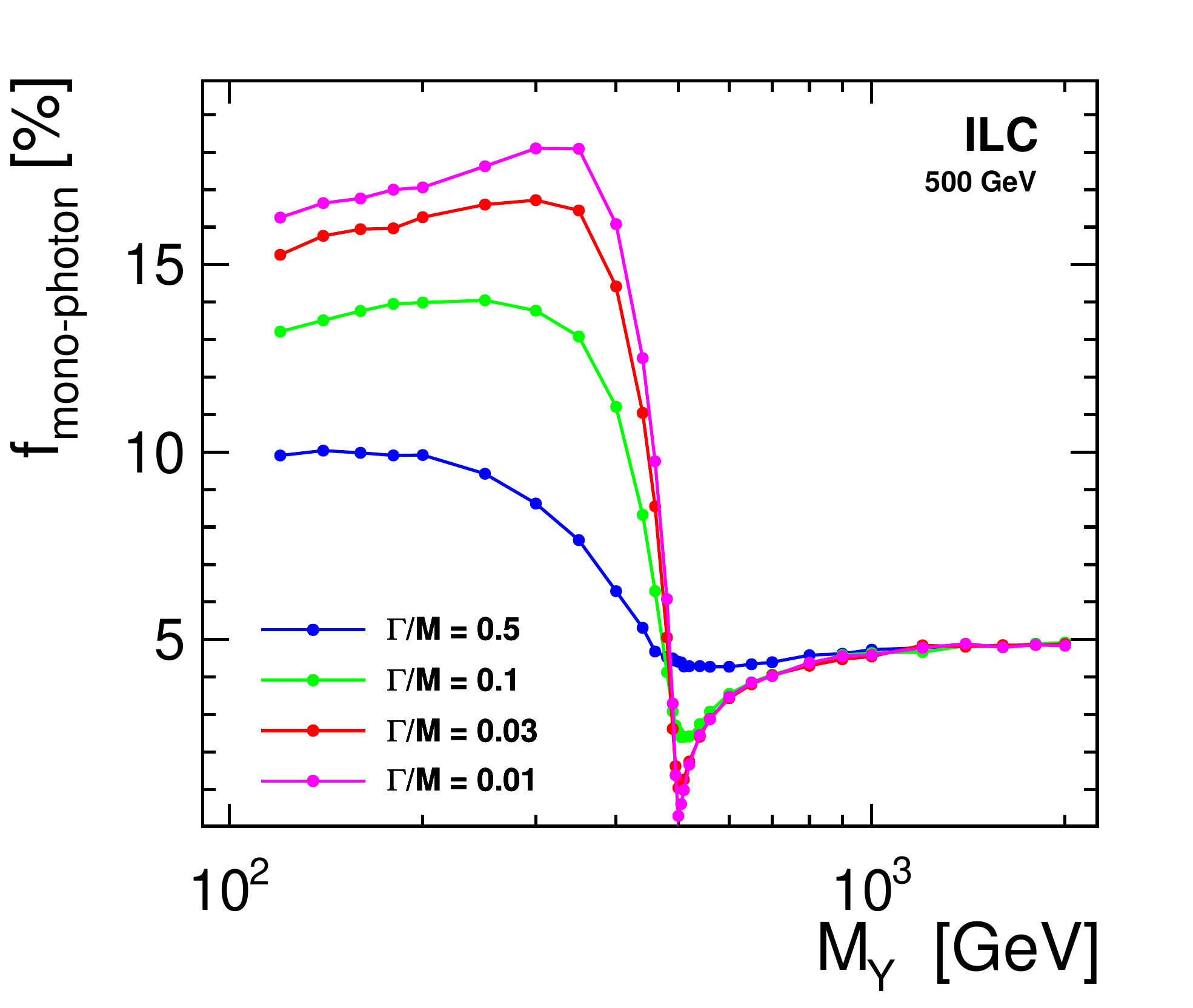}
 \includegraphics[width=0.49\textwidth]{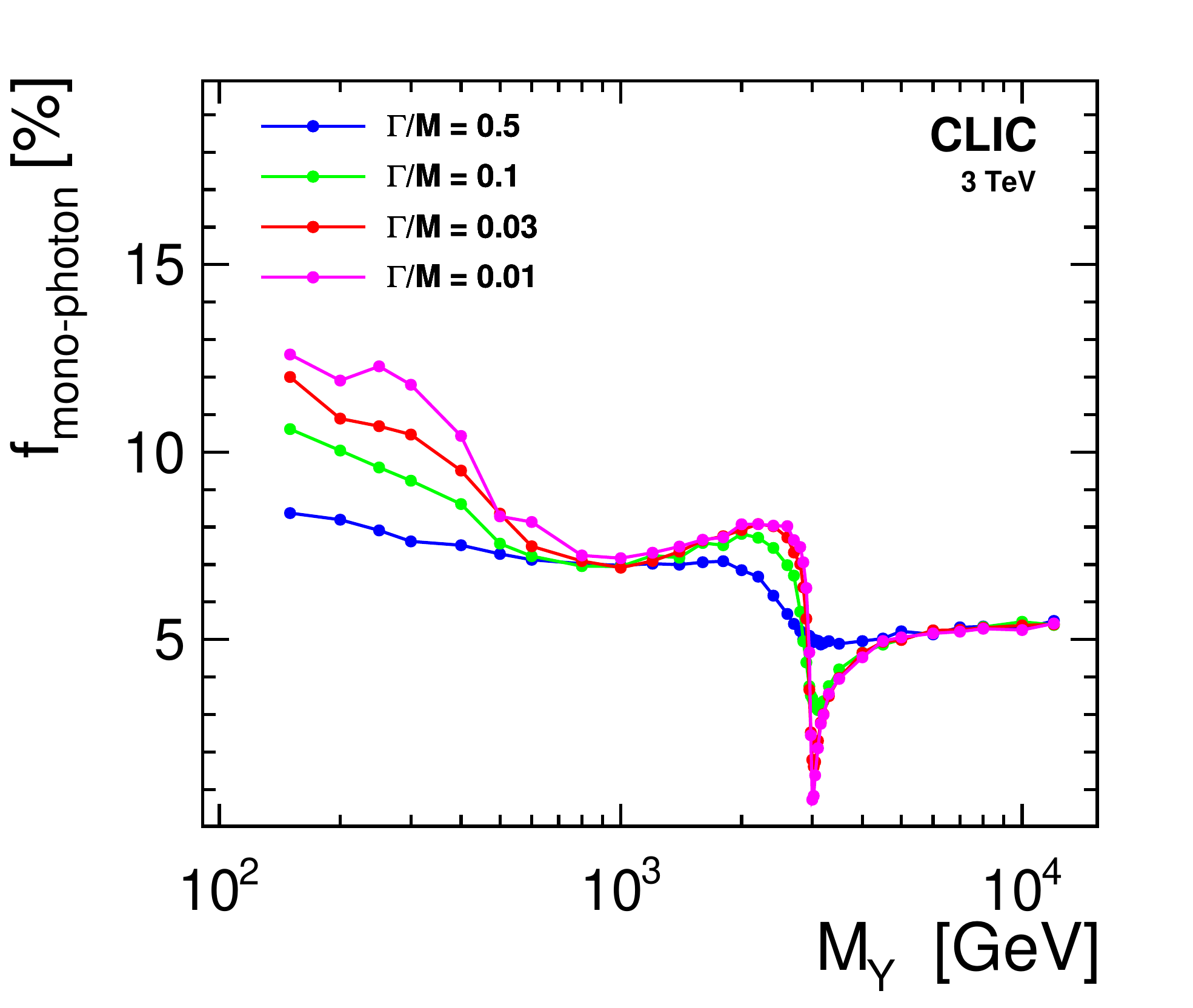}
  \caption{Fraction of dark matter pair-production events, which are
    reconstructed as mono-photon events in the detector, as a function
    of the assumed mediator mass, for the ILC running at 500\,GeV
    (left) and CLIC running at 3\,TeV (right) and different fractional
    mediator widths, as indicated in the plot.
  } 
  \label{fig:det_effi}
\end{figure}

As already mentioned above, both non-radiative and radiative processes
were included in the signal sample generation.
While about half of events included photon generated at the ME level,
only a small fraction of those pass the detector acceptance cuts.
In Fig.~\ref{fig:det_effi}, fraction of signal events, which are
reconstructed as mono-photon events in the detector is shown as a
function of the assumed mediator mass.
Signal selection efficiency of 10--15\% is obtained for low mass
scenarios, $M_Y \ll \sqrt{s}$, while for heavy mediator scenarios,
$M_Y \gg \sqrt{s}$, only about 5\% of events can be tagged.
Similarly to what was observed in Fig.~\ref{fig:isr_rej}, signal
selection efficiency is significantly reduced (following the
probability of hard photon radiation) for resonant mediator
production, $M_Y \approx \sqrt{s}$. 

\subsection{Procedure for setting limits}

\begin{figure} [tb]
 \begin{center}
  \includegraphics[width=0.49\textwidth]{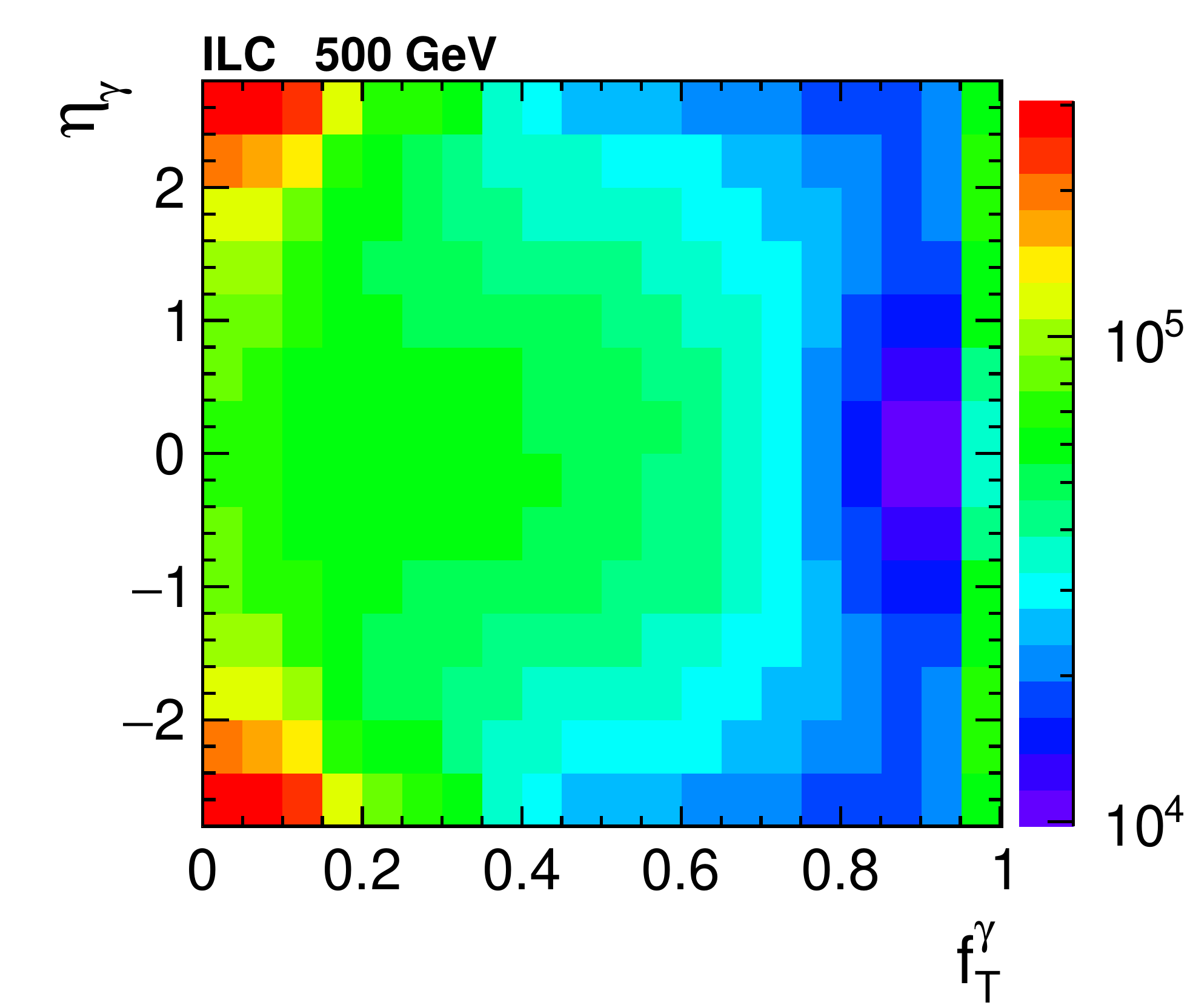}
  \includegraphics[width=0.49\textwidth]{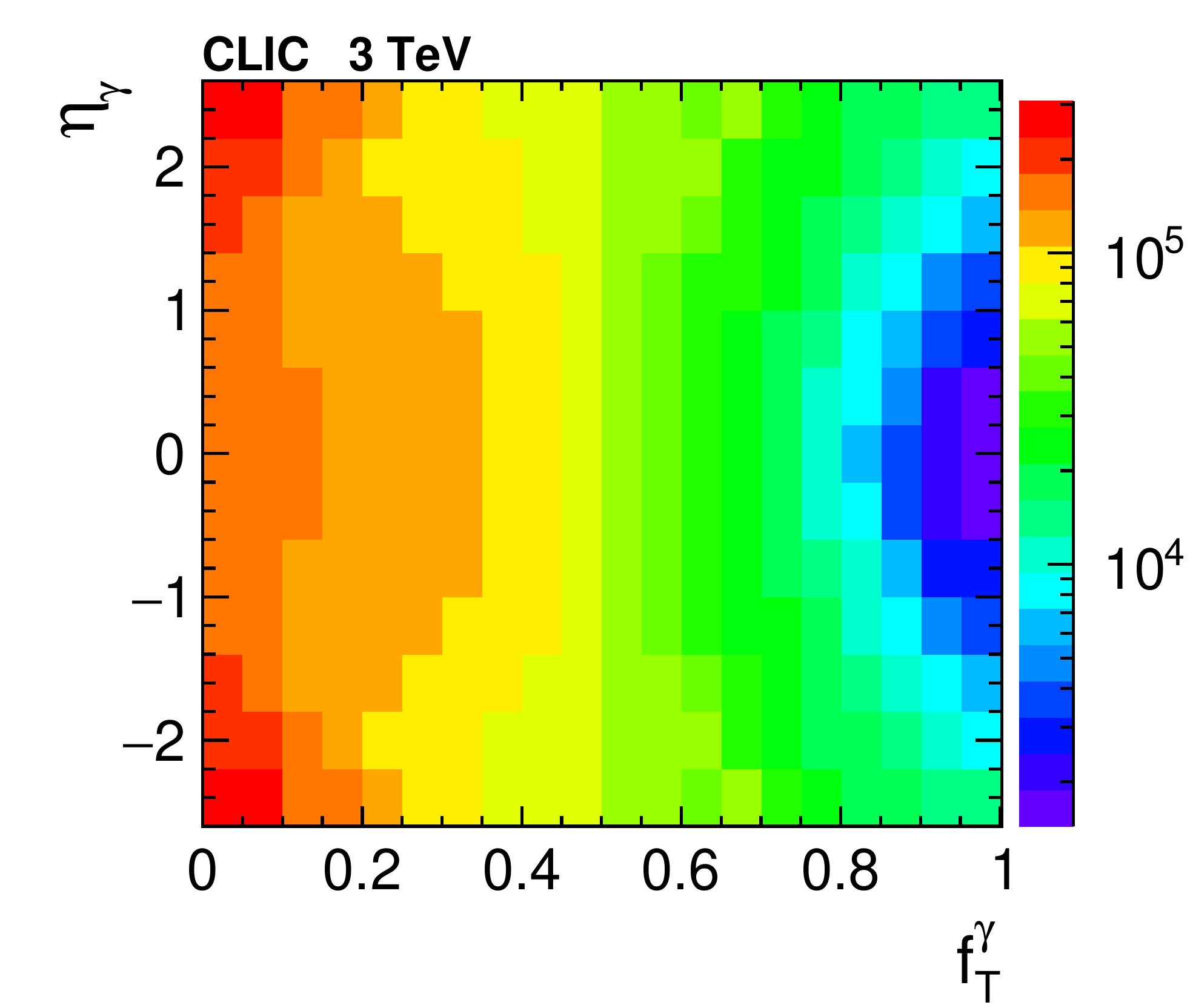}
 \end{center}
\caption{
  Pseudorapidity vs transverse momentum fraction for mono-photon
  background events at 500\,GeV ILC (left) and 3\,TeV CLIC (right).
  Standard model expectations are normalised to integrated luminosity
  of 1600\,fb$^{-1}$ for ILC running with --80\%/+30\% electron/positron
  beam polarisation and 1000\,fb$^{-1}$ for CLIC running with --80\%
  electron beam polarisation only.
}
\label{fig:comp_2d} 
\end{figure}

\begin{figure} [tb]
 \begin{center}
  \includegraphics[width=0.49\textwidth]{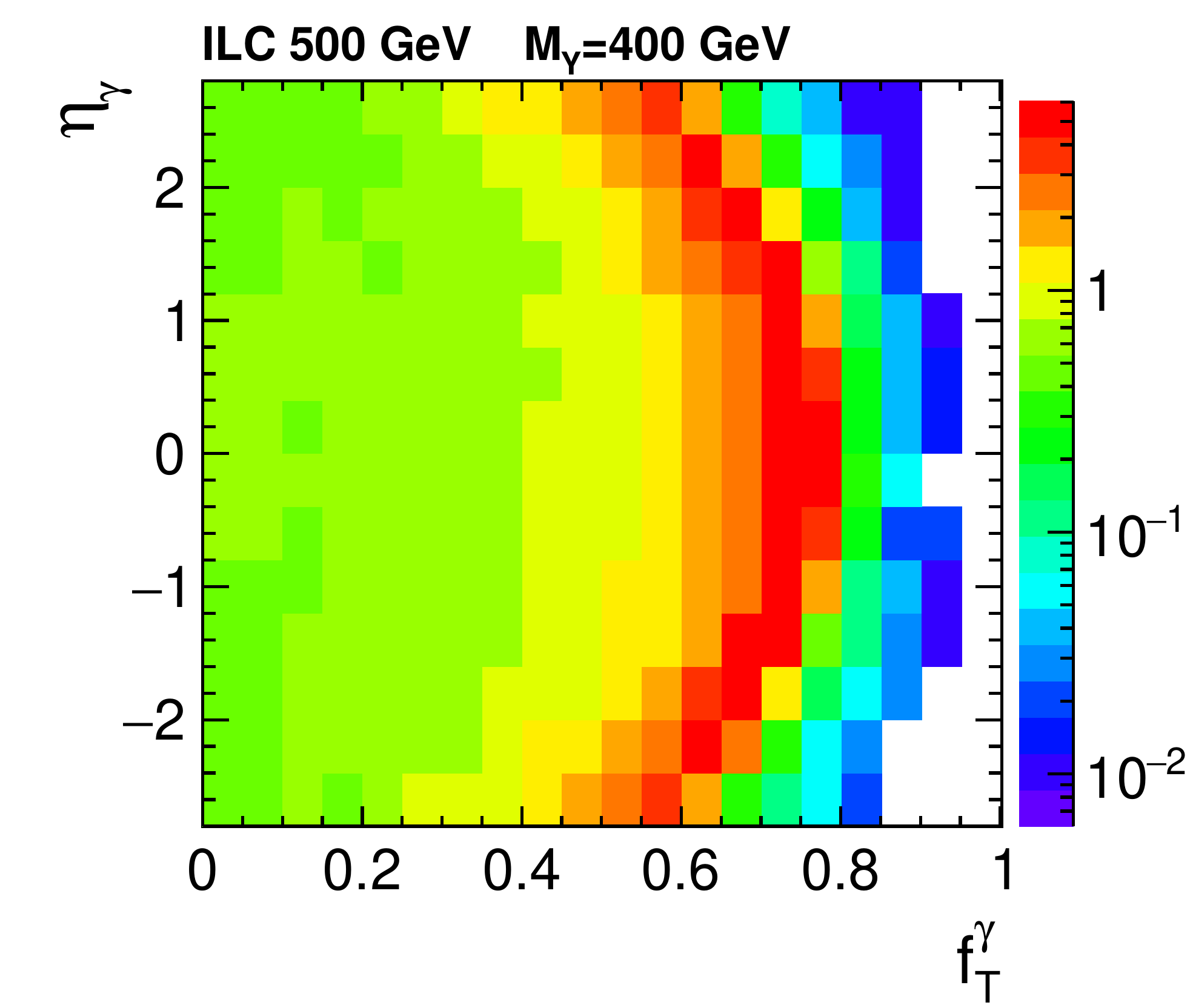}
  \includegraphics[width=0.49\textwidth]{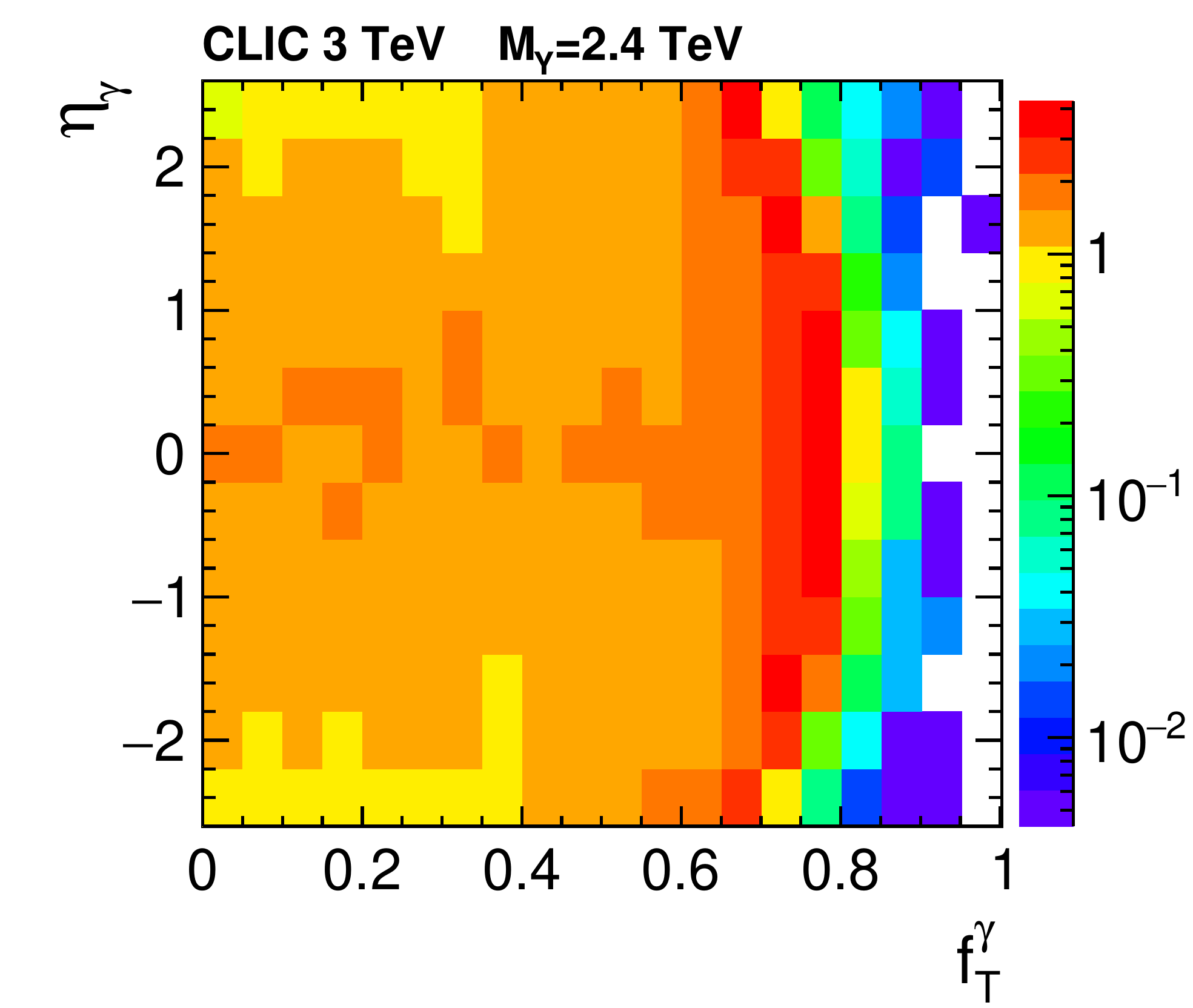}
 \end{center}
\caption{
  Pseudorapidity vs transverse momentum fraction for mono-photon
  signal events at 500\,GeV ILC (left) and 3\,TeV CLIC (right).
  Pair-production of fermion DM particles with $m_\chi$ = 50\,GeV is
  considered for mediator mass $M_Y$ = 400\,GeV at 500\,GeV ILC
  running with --80\%/+30\% electron/positron beam polarisation (left)
  and for $M_Y$ = 2.4\,TeV at 3\,TeV CLIC running with --80\%
  electron beam polarisation only (right).
  Simplified DM model predictions are calculated for the relative
  mediator width of  $\Gamma/M = 0.03$ and normalised to the unpolarised DM production
  cross section of 1\,fb. 
}
\label{fig:sig_2d} 
\end{figure}

Mono-photon events can be described by just two variables: photon
energy and its polar angle or rapidity.
We consider 2-D kinematic distributions of mono-photon events in
rapidity and transverse momentum fraction \ft defined as
\[ \ft = \frac{\log \left( \frac{p_T^{\gamma ~~}}{p_T^{min}} \right)}%
   {\log \left( \frac{p_T^{max}}{p_T^{min}} \right)} \, ,  \] 
where $p_T^{min}$ is the minimum photon transverse momentum required
in the event selection procedure (3\,GeV at 500\,GeV ILC and 10\,GeV
at 3\,TeV CLIC; see Sec.~\ref{sec:selection}) and $p_T^{max}$ is the
maximum transverse momentum of the photon allowed for the given
scattering angle 
\[ p_T^{max} = \frac{\sqrt{s}}{2} \sin \theta_\gamma  \, . \]
Such a choice of variables results in a rectangular phase space region
covered by the accepted mono-photon events and reduces the impact of
statistical fluctuations.
Shown in Fig.~\ref{fig:comp_2d} is the rapidity vs transverse momentum
fraction distributions for mono-photon events expected
for SM background processes at 500\,GeV ILC and 3\,TeV CLIC.
Background from radiative Bhabha events remaining after mono-photon
selection contributes mainly to the region of low \ft (low transverse
momentum) and large photon rapidities while radiative neutrino
production dominates the large \ft region.
The visible difference in the expected background shape for the ILC and CLIC
is mainly due to the fact that the cross section for radiative Bhabha
scattering decreases fast with energy and the radiative neutrino
pair production cross section increases \cite{Kalinowski:2020lhp}.
Corresponding distributions for examples of
signal scenarios are shown in Fig.~\ref{fig:sig_2d}. 
The shape of the expected signal distribution is very different from
that for the SM background.
There is a clear ridge visible, corresponding to radiative
return events, when the mediator is produced on-shell.  
We use these 2-D distributions as an input for building measurement
model in RooFit \cite{Verkerke:2003ir} and calculate the expected  
95\% C.L. cross section limits for radiative DM pair-production
in the CL$_\textrm{S}$ approach \cite{Read:2002hq}.

\subsection{Impact of beam polarisation}

\begin{figure}[tbp]
 \includegraphics[width=0.49\textwidth]{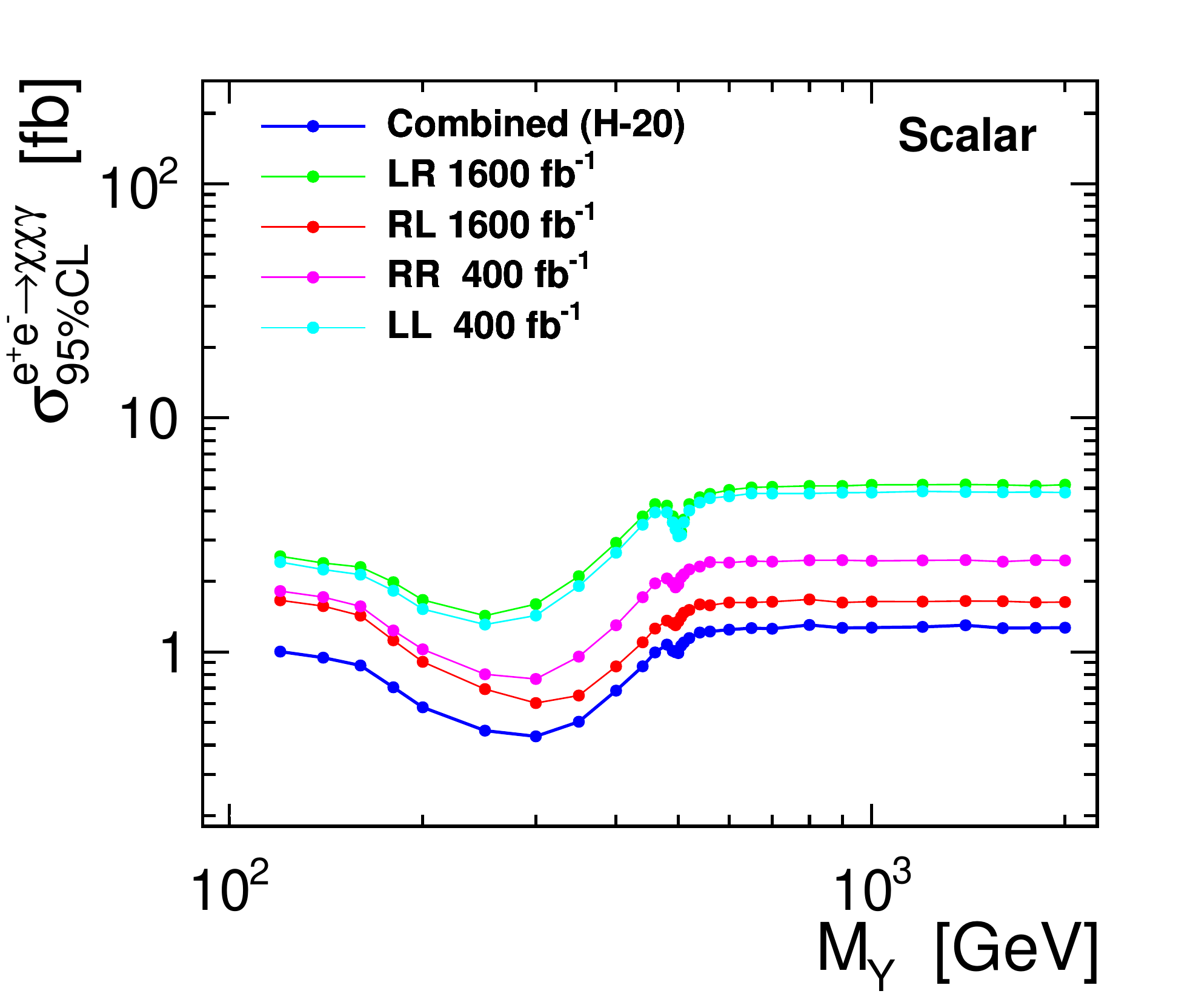}
 \includegraphics[width=0.49\textwidth]{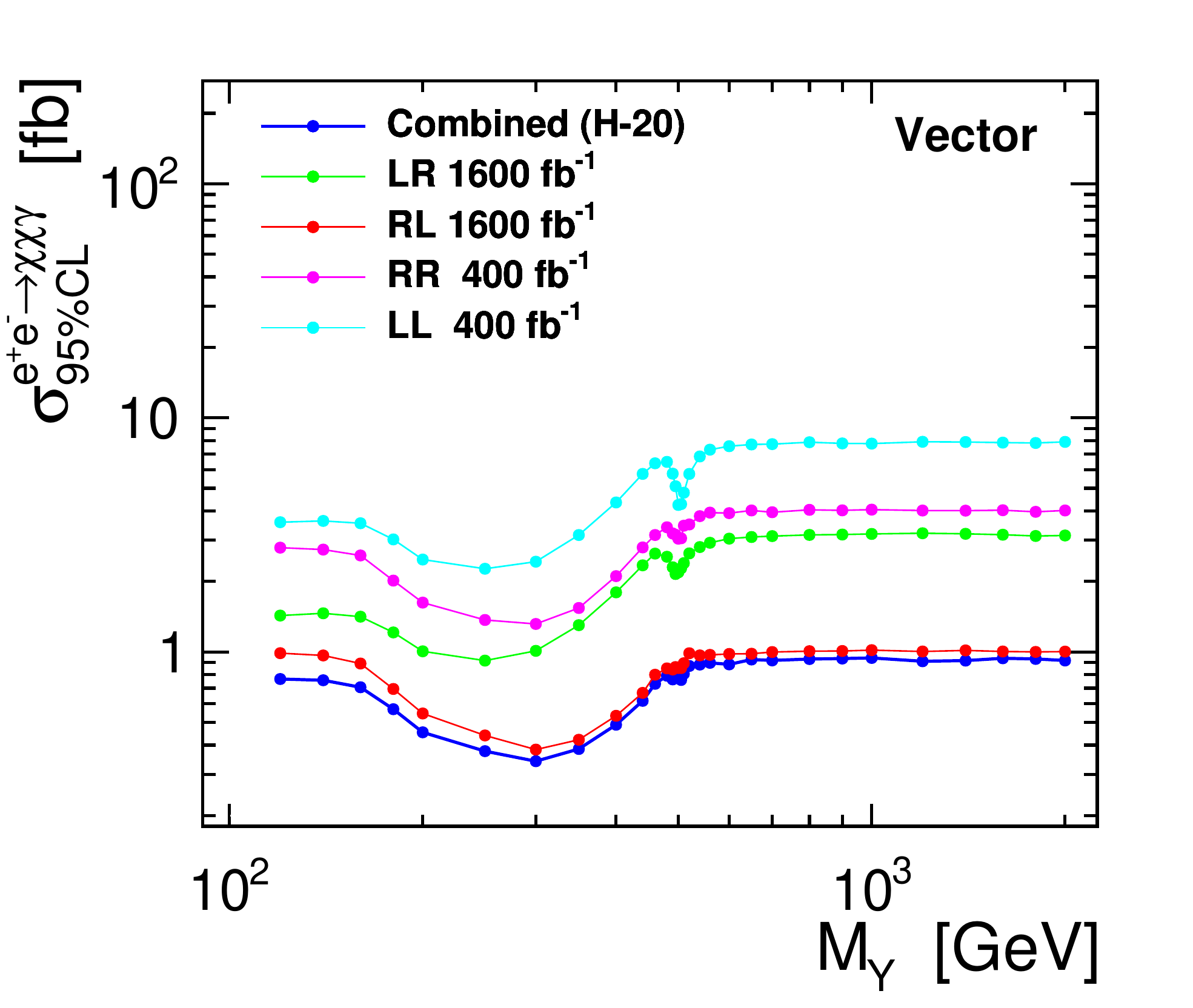}
 \includegraphics[width=0.49\textwidth]{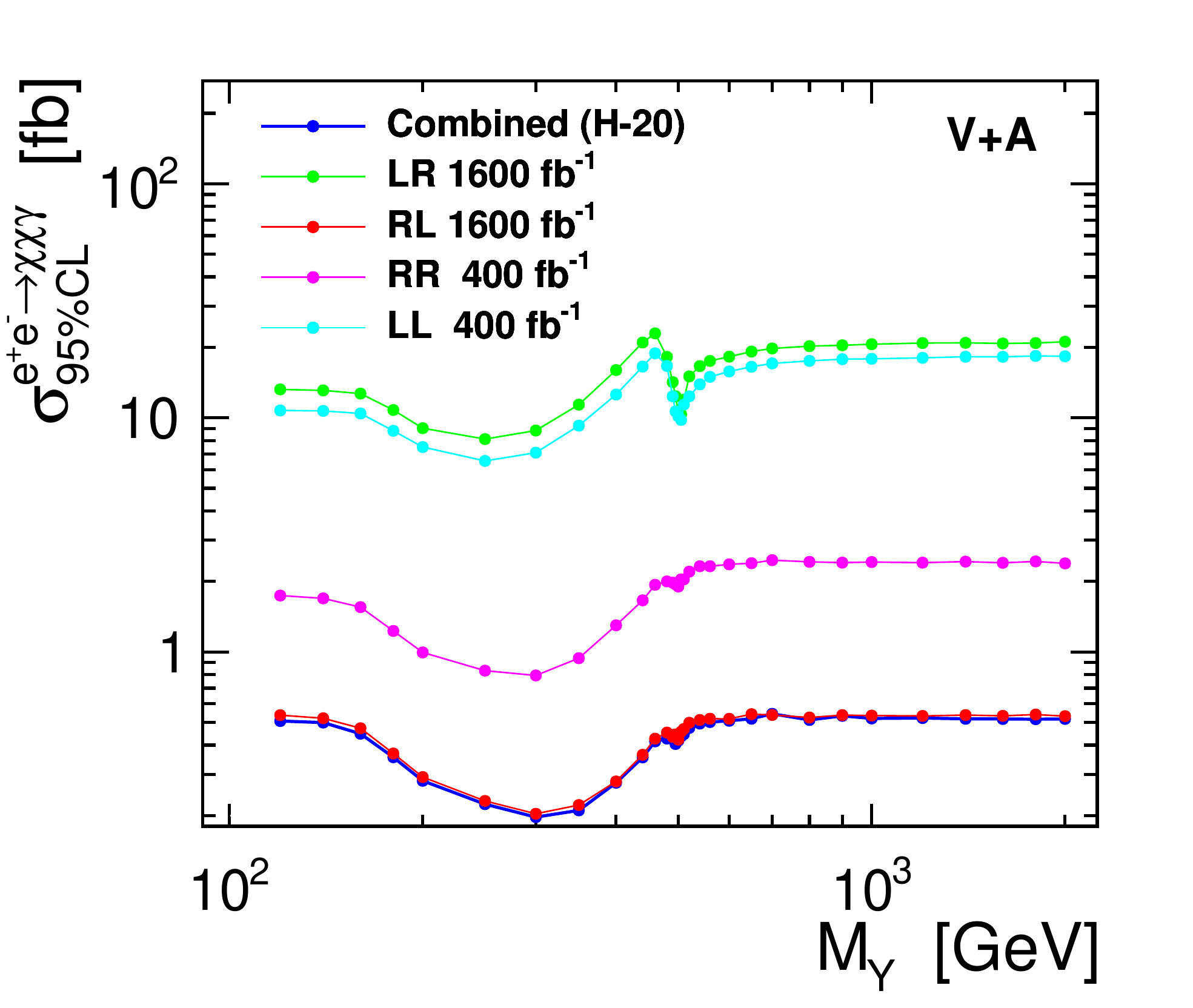}
 \includegraphics[width=0.49\textwidth]{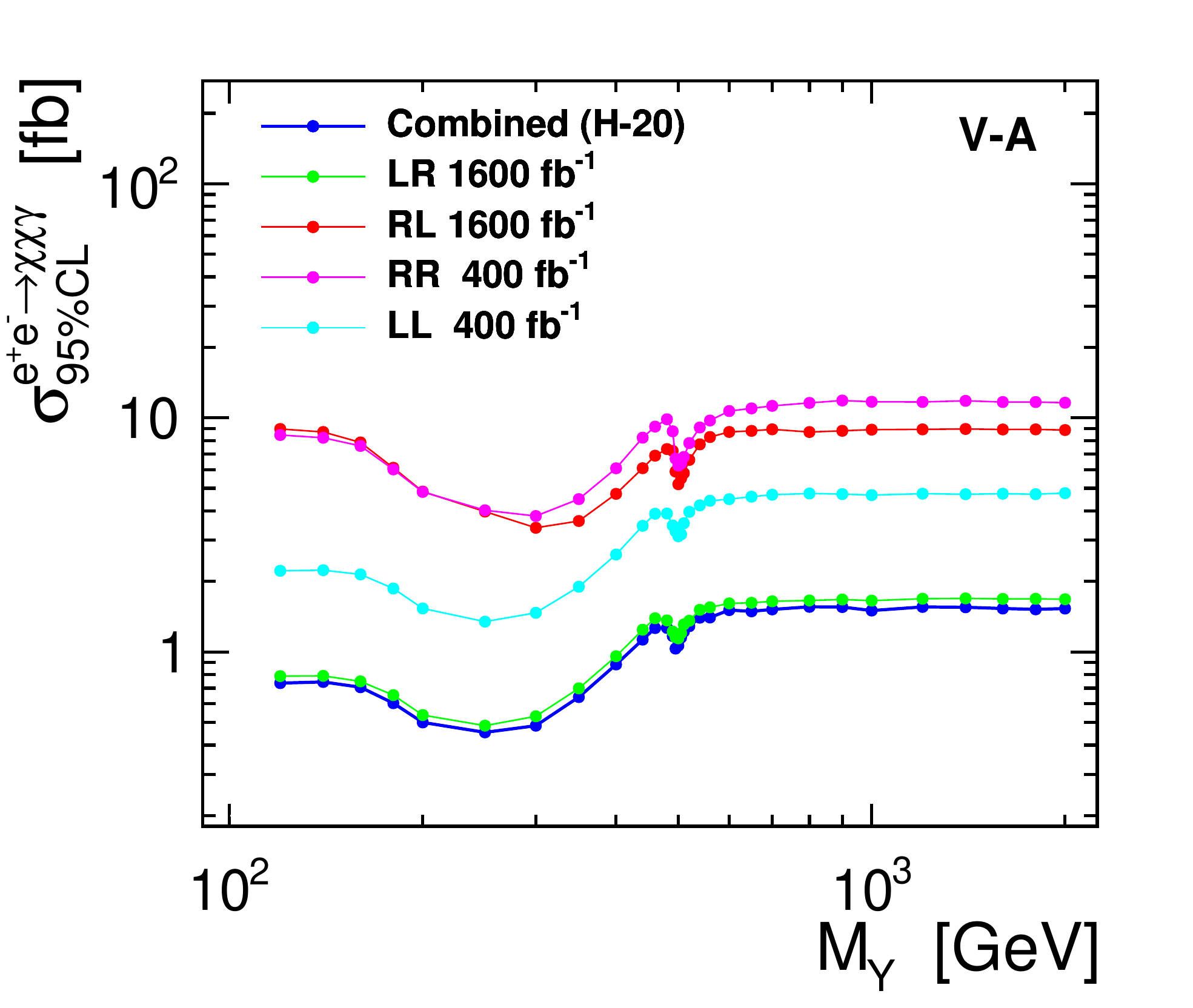}
 \caption{Limits on the cross section for the radiative light DM
   pair-production processes with $s$-channel mediator exchange
   for  500\,GeV ILC. Different mediator coupling scenarios are
   considered, as indicated in the plot label (see
   Tab.~\ref{tab:model_scenarios} for details). 
   Limits resulting from running with different beam polarisations and
   the combined limits are presented.
    Systematic uncertainties are not taken into account.
  } 
  \label{fig:res_pol}
  
\end{figure}

While the final results from the analysis will be presented only for
combined analysis of all runs,
taken with different beam polarisation settings,
it is important to understand the impact
of the beam polarisation on the experimental sensitivity.
Shown in Fig.~\ref{fig:res_pol} are the expected limits on the
radiative light DM pair-production cross section at 500\,GeV ILC for
four\footnote{
Six scenarios are defined in  Tab.~\ref{tab:model_scenarios}. However,
polarisation dependences are not shown for Pseudo-scalar and
Pseudo-vector scenarios as they are very similar to
those observed for Scalar and Vector scenarios, respectively.
}
mediator scenarios and different polarisation configurations
considered. 
Running with right-handed electron beam polarisation and
left-handed positrons (RL configuration) results in smallest
background from neutrino radiative pair-production and thus also the
strongest limits for most of the considered scenarios: Scalar, Vector
and V+A coupling structure.
However, for scenario with V--A coupling structure data collected with
LR polarisation configuration is expected to provide much stronger limits. 
This demonstrates that the optimal choice of the beam polarisation
depends on the assumed mediator coupling structure and combined
analysis of the data collected for all polarisation combinations is
required to set most stringent limits for all considered coupling
scenarios. 
Same conclusion holds also for the CLIC running scenarios.

\subsection{Systematic uncertainties}

With high numbers of the expected background events, see
Fig.~\ref{fig:comp_2d}, statistical uncertainties of the measurement
are expected to be on a sub-percent level.
Therefore, possible sources of systematic uncertainties have to be
taken into account.
When different data sets are combined, it is
also very important to take possible correlations in
systematic deviations into account.
Systematic variations and their correlations were included in the definition of
the measurement model in the RooFit framework.

In our choice of systematic effects we follow the approaches 
adopted in \cite{Blaising:2021vhh,Habermehl:2020njb}. 
The following sources of systematic uncertainties were considered in the
presented study:
\begin{itemize}
\item neutrino background normalisation;\\
  We assume that the SM predictions for the radiative neutrino
  pair-production background normalisation are known with relative
  uncertainty of 0.2\%  \cite{Blaising:2021vhh} (for both ILC and CLIC).
  This includes the uncertainty in theoretical predictions as well as 
  the possible uncertainty in the event selection efficiency. The
  uncertainty is 100\% correlated between different data sets.
\item Bhabha background normalisation;\\
  Most of Bhabha background events are removed by selection cuts,
  vetoes applied based on BeamCal and LumiCal response in particular.
  High rejection efficiency results in an increased relative
  uncertainty at the level of accepted events.
  We assume that the predictions for the Bhabha event background are
  known to 1\%  \cite{Blaising:2021vhh} (for both ILC and CLIC).
  This includes the uncertainty in theoretical predictions as well.
  The uncertainty is 100\% correlated between different data sets.
\item integrated luminosity of the data samples;\\
  We assume 0.26\% luminosity uncertainty for 500\,GeV
  ILC~\cite{BozovicJelisavcic:2013lni} and 0.2\% uncertainty for 3\,TeV
  CLIC~\cite{Lukic:2013fw}. We consider this uncertainty as 
  uncorrelated between different data sets.
\item beam polarisation; \\
  We assume relative uncertainty on the beam polarisation at ILC of
  0.02--0.08\%, following approach of \cite{Karl:2017xra}.
  Uncertainties are correlated between runs with the same electron or
  positron beam polarisation. For CLIC, uncertainty of 0.2\% is
  assumed for the electron beam polarisation \cite{Wilson:lcws2012}
  (uncorrelated).
\item shape of the luminosity spectra;\\
  We consider possible variation of the luminosity spectra shape by
  reweighting the signal and background events accordingly. Results of
  \cite{Habermehl:2020njb} were used for ILC, with systematic variation
  of spectra normalisation of up to 50\% at half of the nominal
  centre-of-mass energy ($\sqrt{s'} = \frac{1}{2}\sqrt{s}$).
  We use the same approach for spectra shape uncertainty at CLIC.
  However, as the low energy tail of the luminosity spectra is much
  higher at CLIC than at ILC, we assume that its relative uncertainty
  is an order of magnitude lower and corresponds to 5\% normalisation
  uncertainty at half of the nominal centre-of-mass energy. This
  uncertainty is 100\% correlated between different data sets.
\end{itemize}

\begin{figure}[tbp]
 \includegraphics[width=0.49\textwidth]{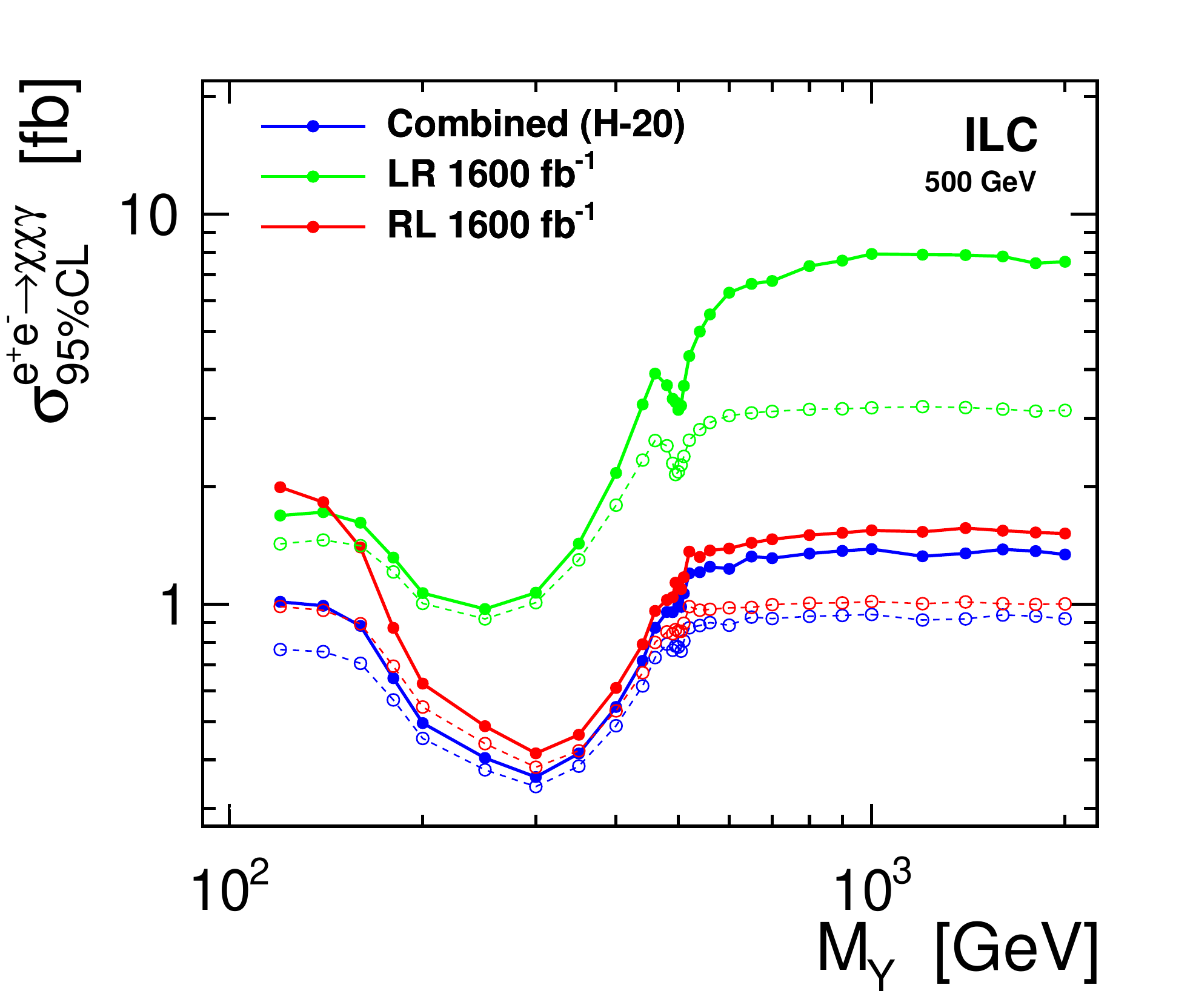}
 \includegraphics[width=0.49\textwidth]{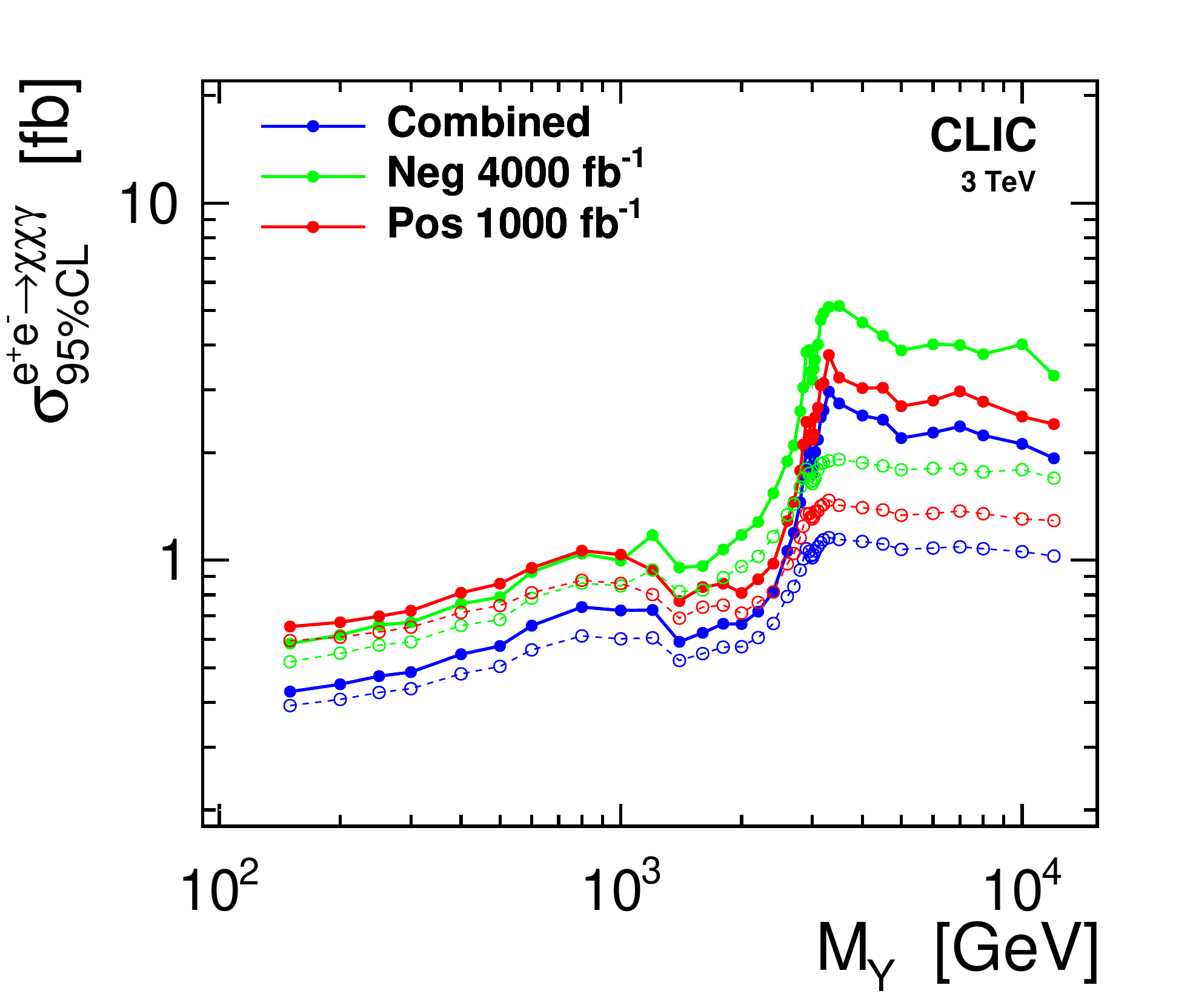}
 \includegraphics[width=0.49\textwidth]{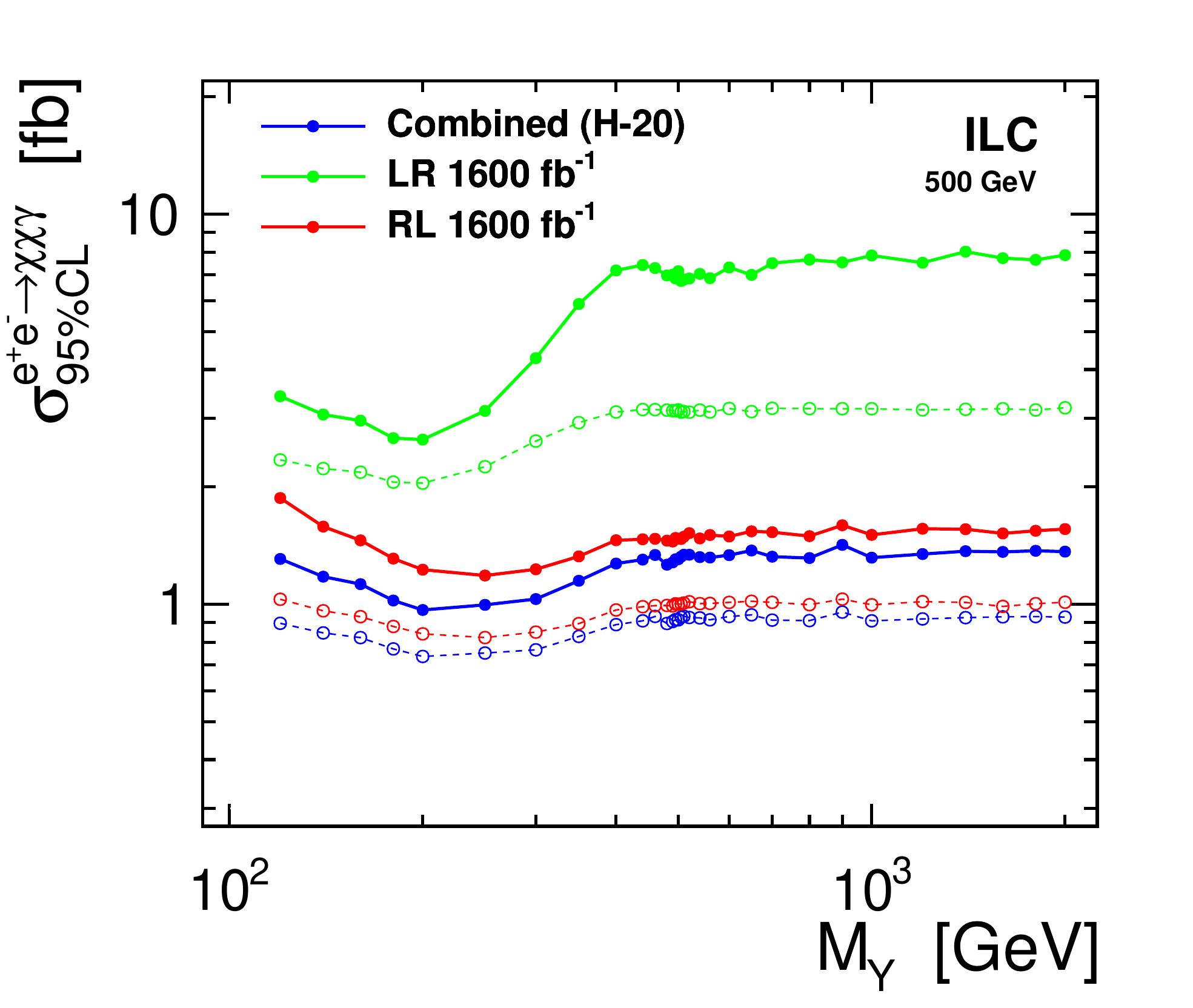}
 \includegraphics[width=0.49\textwidth]{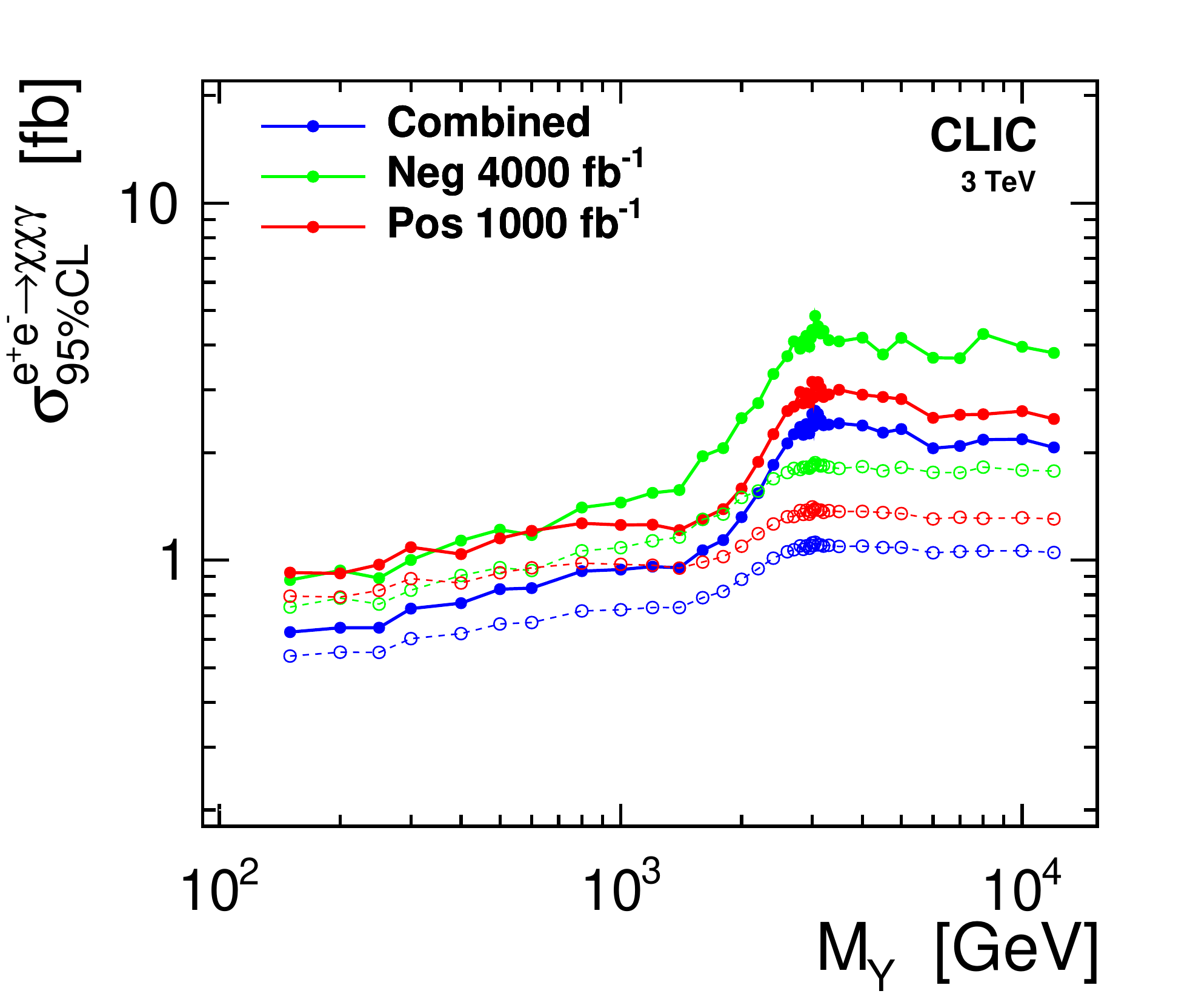}
  \caption{Limits on the cross section for the radiative light DM
    pair-production processes with vector mediator exchange at
    500\,GeV ILC (left) and 3\,TeV CLIC (right). Limits extracted with
    (solid line) and without (dashed line) taking into account
    systematic uncertainties are compared for narrow mediator
    ($\Gamma/M = 0.03$; top row) and wide mediator ($\Gamma/M = 0.5$;
    bottom) hypotheses.
  } 
  \label{fig:res_sys}
  
\end{figure}

The influence of the systematic uncertainties on the analysis results
is illustrated in Fig.~\ref{fig:res_sys}. 
Expected limits on the radiative DM pair-production cross section with
vector mediator exchange are presented for  500\,GeV ILC and 3\,TeV
CLIC without and with systematic uncertainties taken into account.
Impact of systematic uncertainties is most significant for heavy
mediator exchange, for $M_Y > \sqrt{s}$.
When systematic effects are taken into account, expected CLIC combined
limits increase by about a factor of 2, with largest contribution
coming from the polarisation uncertainty. 
For ILC, the systematic uncertainties increase the expected combined
limits for heavy mediator scenarios by about 40\%, with largest
contributions coming from the integrated luminosity and spectra shape
uncertainties. 
%


\section{Results}
\label{sec:results}

\begin{figure}[tbp]
 \includegraphics[width=0.49\textwidth]{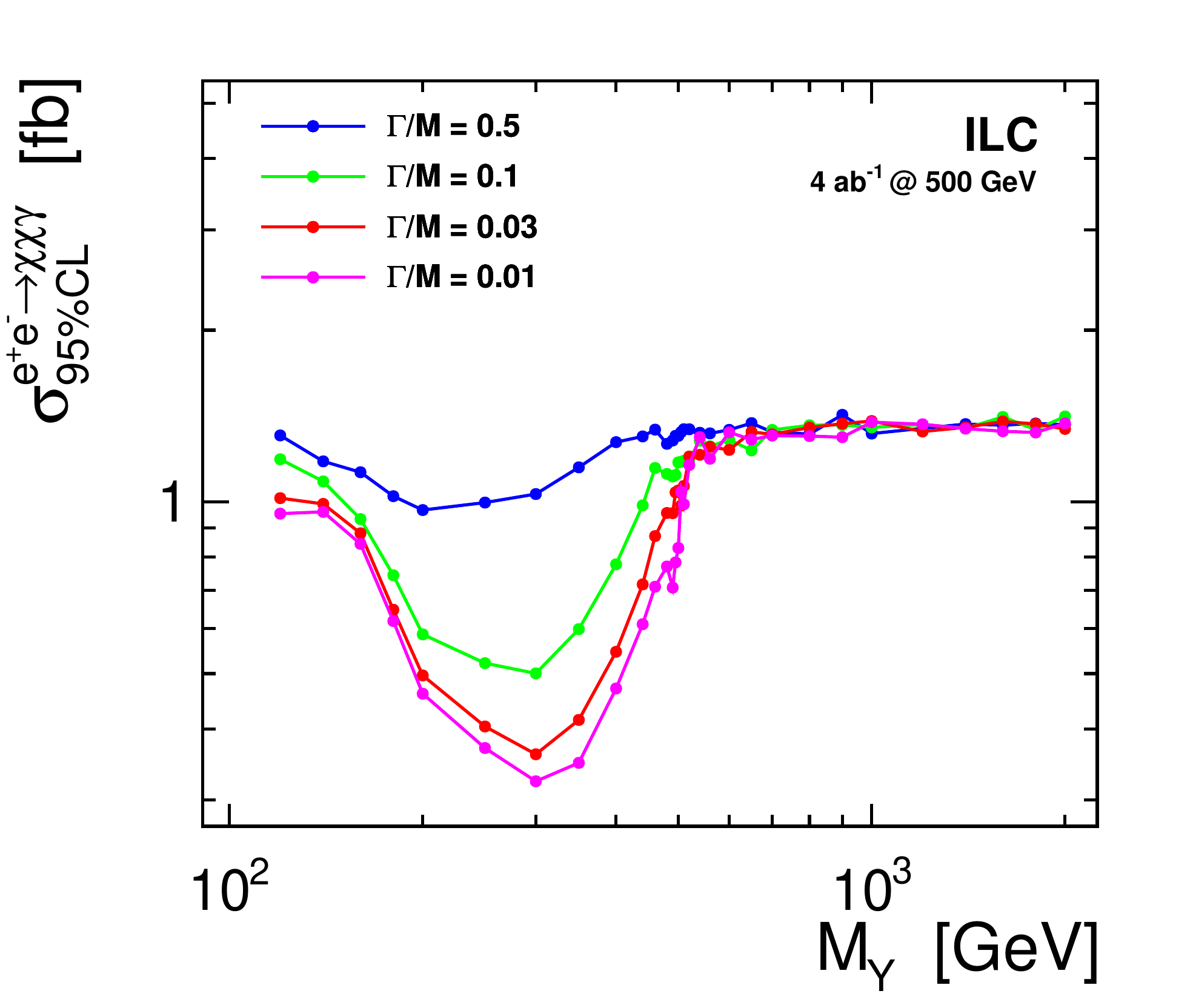}
 \includegraphics[width=0.49\textwidth]{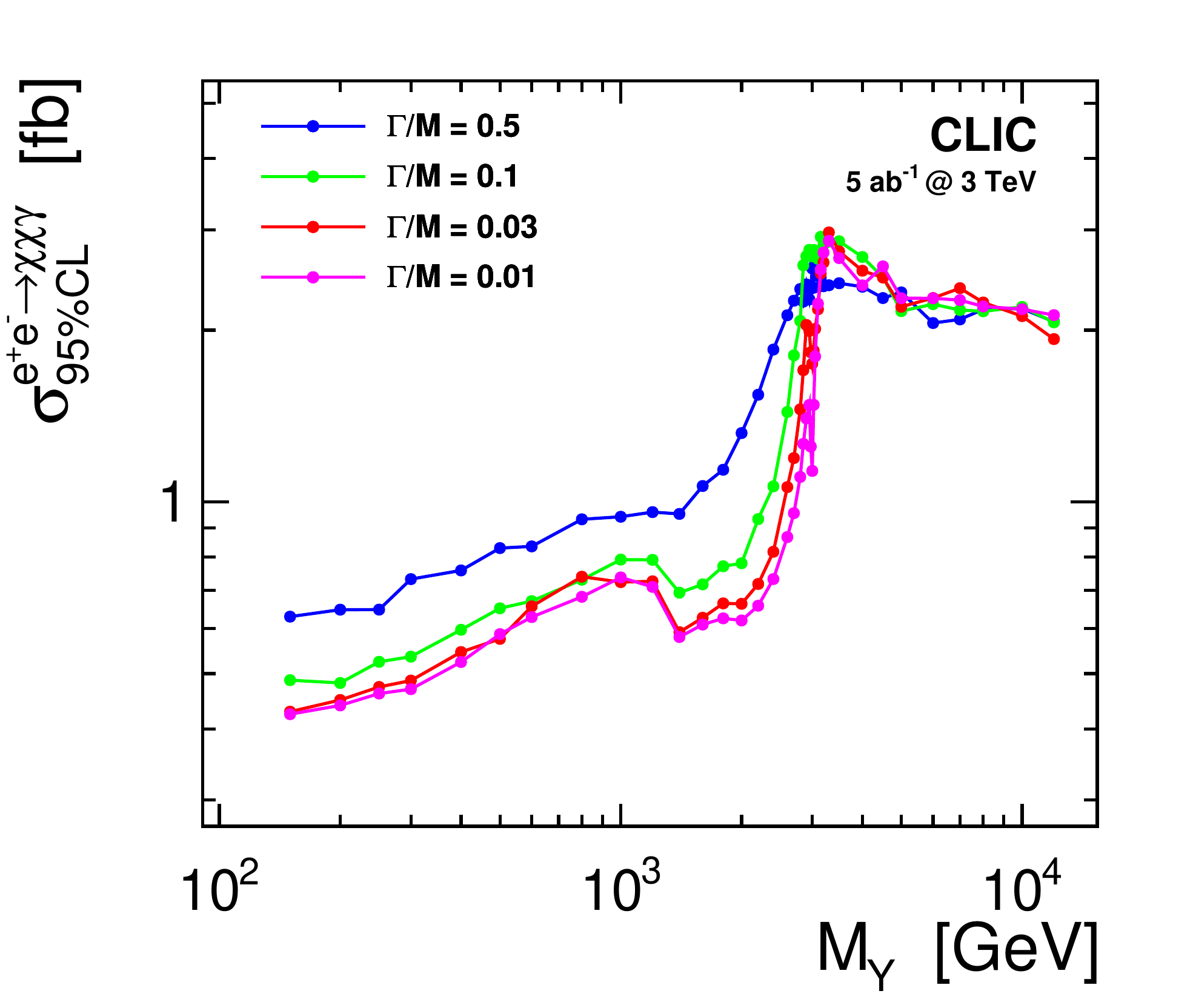}
  \caption{Limits on the cross section for the radiative DM
pair-production processes with $s$-channel  vector mediator exchange
for the ILC running at 500\,GeV (left) and CLIC running at 3\,TeV
(right) and different fractional  mediator widths, as indicated in the
plot. Combined limits corresponding to the assumed running scenarios
are presented with systematic uncertainties taken into account.
  } 
  \label{fig:res2_wid}
\end{figure}

Presented in Fig.~\ref{fig:res2_wid} are the expected limits on the
cross section for radiative DM pair-production at 500\,GeV ILC and
3\,TeV CLIC, assuming combined analysis of all collected data, as
described in sec.~\ref{sec:running}.
Limits are presented for the Vector mediator scenario as a function of
the mediator mass for different  mediator widths, for the Dirac
fermion DM with mass $m_\chi = 50$\,GeV.
For heavy mediator scenarios, $M_{Y} \gg \sqrt{s}$, expected cross
section limits do not depend on the assumed mediator width and weakly depend
on the assumed mediator mass.
For lower masses, the cross section limits improve and the expected
sensitivity is highest, as expected, for narrow mediator scenarios.
Surprisingly, the shape of the mass dependence below the resonant
production threshold is very different for ILC and CLIC.
This can be understood as an effect of the much wider luminosity
spectra at CLIC significantly enhancing the resonant production cross
section for $M_{Y} < \sqrt{s}$, as already observed in Fig.~\ref{fig:exp_mass}.
For ILC running at 500\,GeV, the best limits are obtained for resonance with 
$M_{Y} \approx \frac{1}{2}\sqrt{s}$, while lowest mediator masses are
constrained best for CLIC.
Still, one should note that the presented 95\% C.L. limits on the
radiative DM pair-production depend rather weakly on the mediator mass
and width, while they change by more than an order of magnitude.
The limits are of the order of 1\,fb, ranging from about
0.3\,fb to 3\,fb.

\begin{figure}[tbp]
 \includegraphics[width=0.49\textwidth]{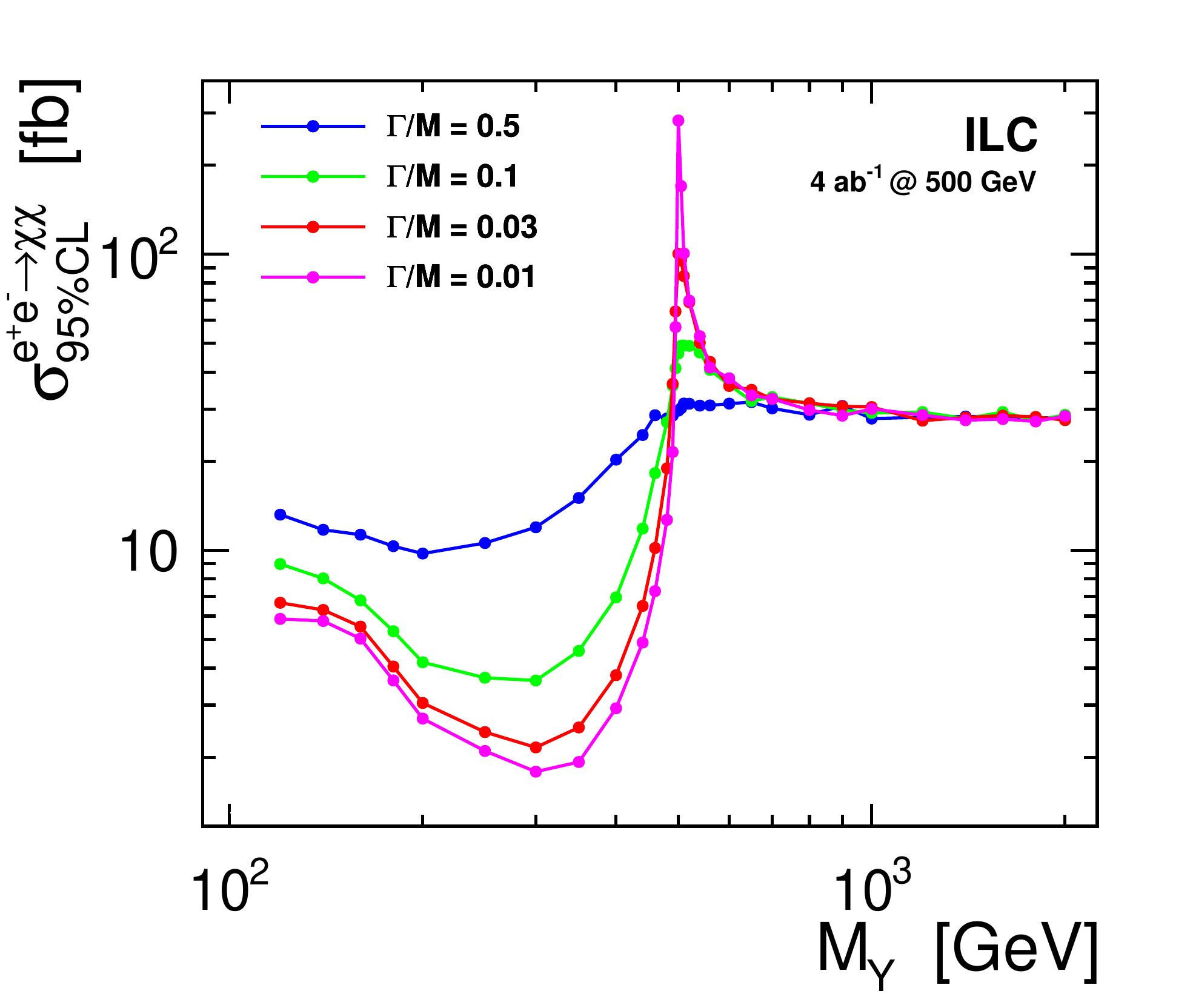}
 \includegraphics[width=0.49\textwidth]{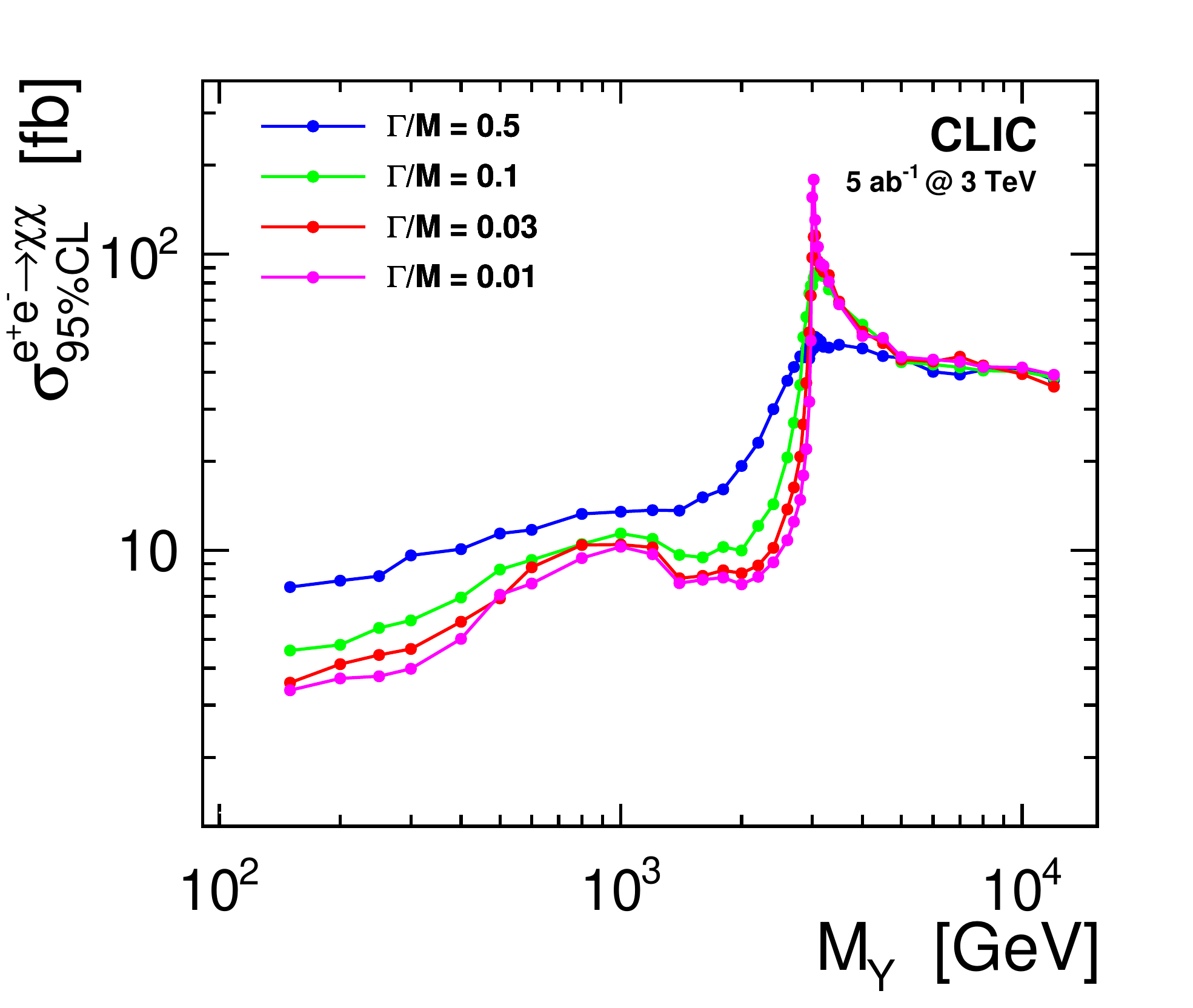}
  \caption{Limits on the cross section for light fermionic DM
pair-production processes with $s$-channel  vector mediator exchange
for the ILC running at 500\,GeV (left) and CLIC running at 3\,TeV
(right) and different fractional  mediator widths, as indicated in the
plot. Combined limits corresponding to the assumed running scenarios
are presented with systematic uncertainties taken into account.
  } 
  \label{fig:res_wid}
\end{figure}

Limits presented in Fig.~\ref{fig:res2_wid} correspond to processes
with hard photon reconstructed in the detector
and are therefore sensitive to detector design and analysis details
(e.g. the photon transverse momentum or energy thresholds applied).
To be able to constrain different BSM scenarios, the limits need to be
corrected for the photon tagging efficiency, as discussed in
sec.~\ref{sec:selection} and shown in Fig.~\ref{fig:det_effi}.
Resulting limits on the total DM pair-production cross section
at 500\,GeV ILC and 3\,TeV CLIC are presented in Fig.~\ref{fig:res_wid}. 
As expected, sensitivity to models with $M_Y \approx \sqrt{s}$ is
significantly reduced except for the very wide mediator scenario.
In general, expected limits on the DM pair-production with light
mediator exchange are of the order of 10\,fb for both ILC and CLIC.

\begin{figure}[tb]
 \includegraphics[width=0.49\textwidth]{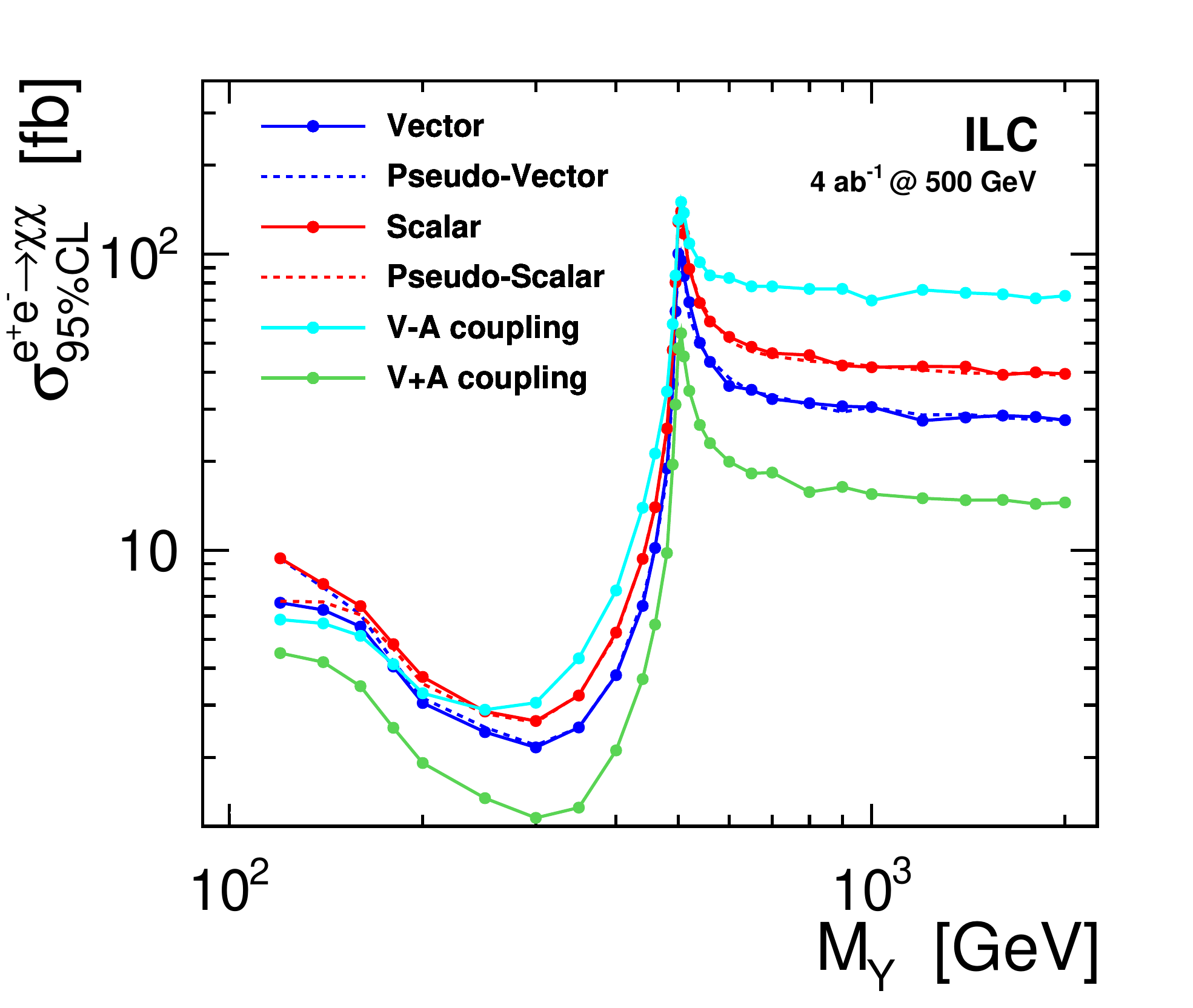}
 \includegraphics[width=0.49\textwidth]{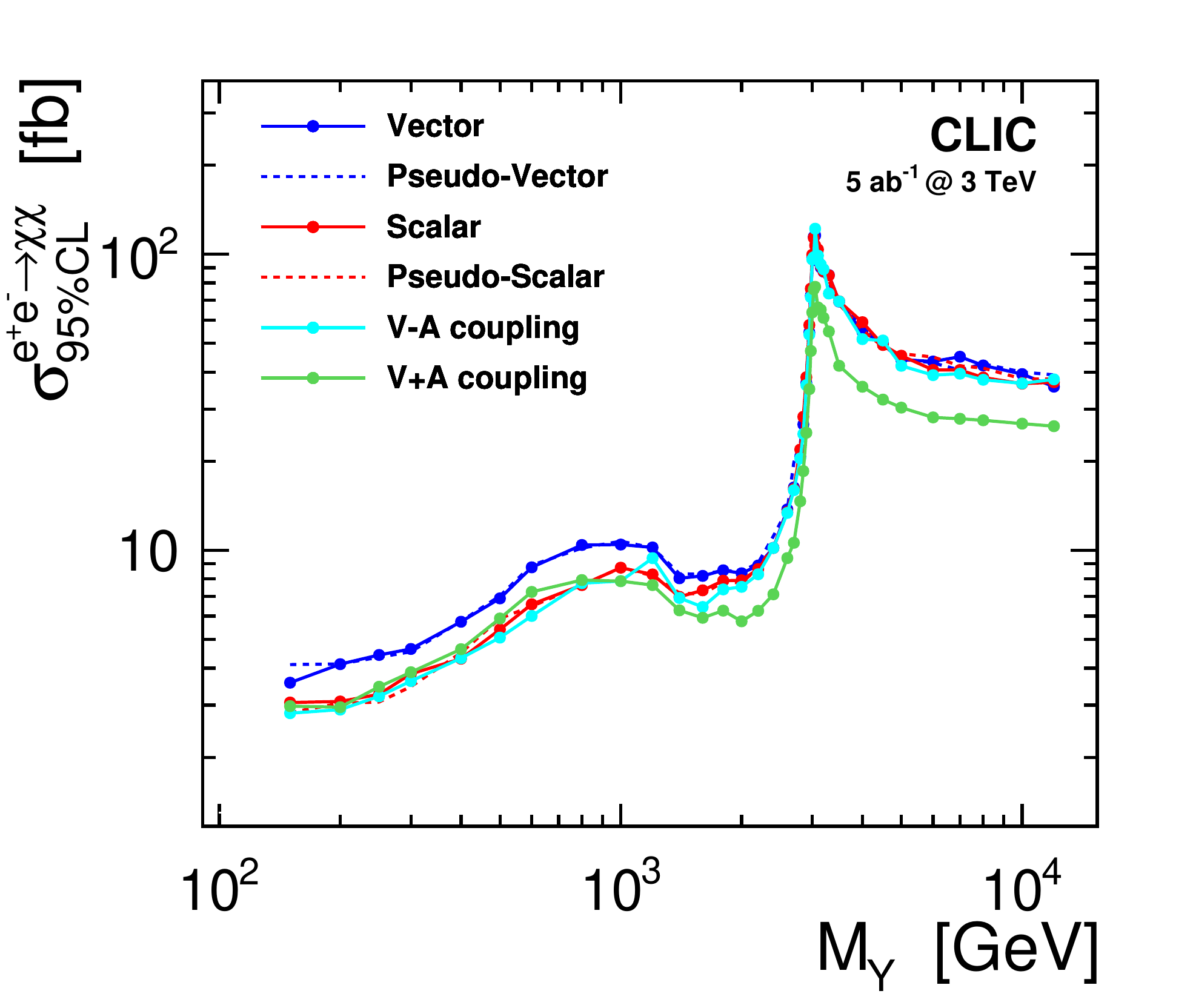}
  \caption{Limits on the cross section for light fermionic DM
pair-production processes with $s$-channel mediator exchange for
the ILC running at 500\,GeV (left) and CLIC running at 3\,TeV
(right), for relative mediator width, $\Gamma/M = 0.03$, and different
mediator coupling scenarios, as indicated in the plot. Combined limits
corresponding to the assumed running scenarios are presented with
systematic uncertainties taken into account.  
  } 
  \label{fig:res_model}
\end{figure}
Compared in Fig.~\ref{fig:res_model} are the expected limits on the
total DM pair-production cross section at 500\,GeV ILC and 3\,TeV CLIC,
for relative mediator width of $\Gamma/M = 0.03$ and different mediator
coupling scenarios, as listed in Tab.~\ref{tab:model_scenarios}. 
It is interesting to note that
for processes with light mediator exchange the model
dependence of the total cross section limits is weaker than for the
heavy mediator case. 
Also, running scenario assumed for CLIC, with 80\% of integrated
luminosity devoted to running with the negative electron beam
polarisation (with higher background from radiative neutrino
pair-production), results in cross section limits less sensitive to
mediator couplings than the ILC running scenario, where same
luminosity is assumed for  LR and RL configurations.

\begin{figure}[tbp]
 \includegraphics[width=0.49\textwidth]{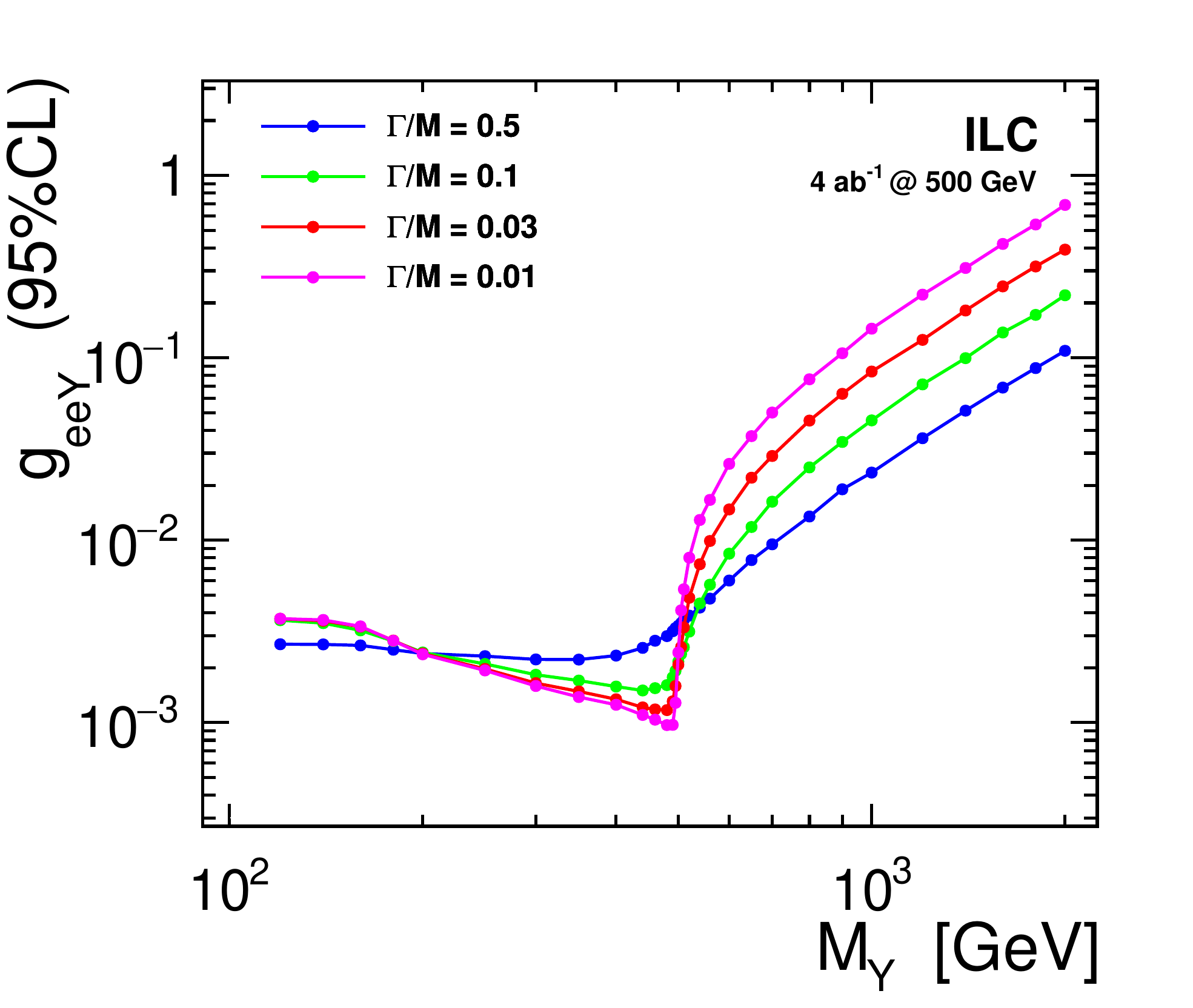}
 \includegraphics[width=0.49\textwidth]{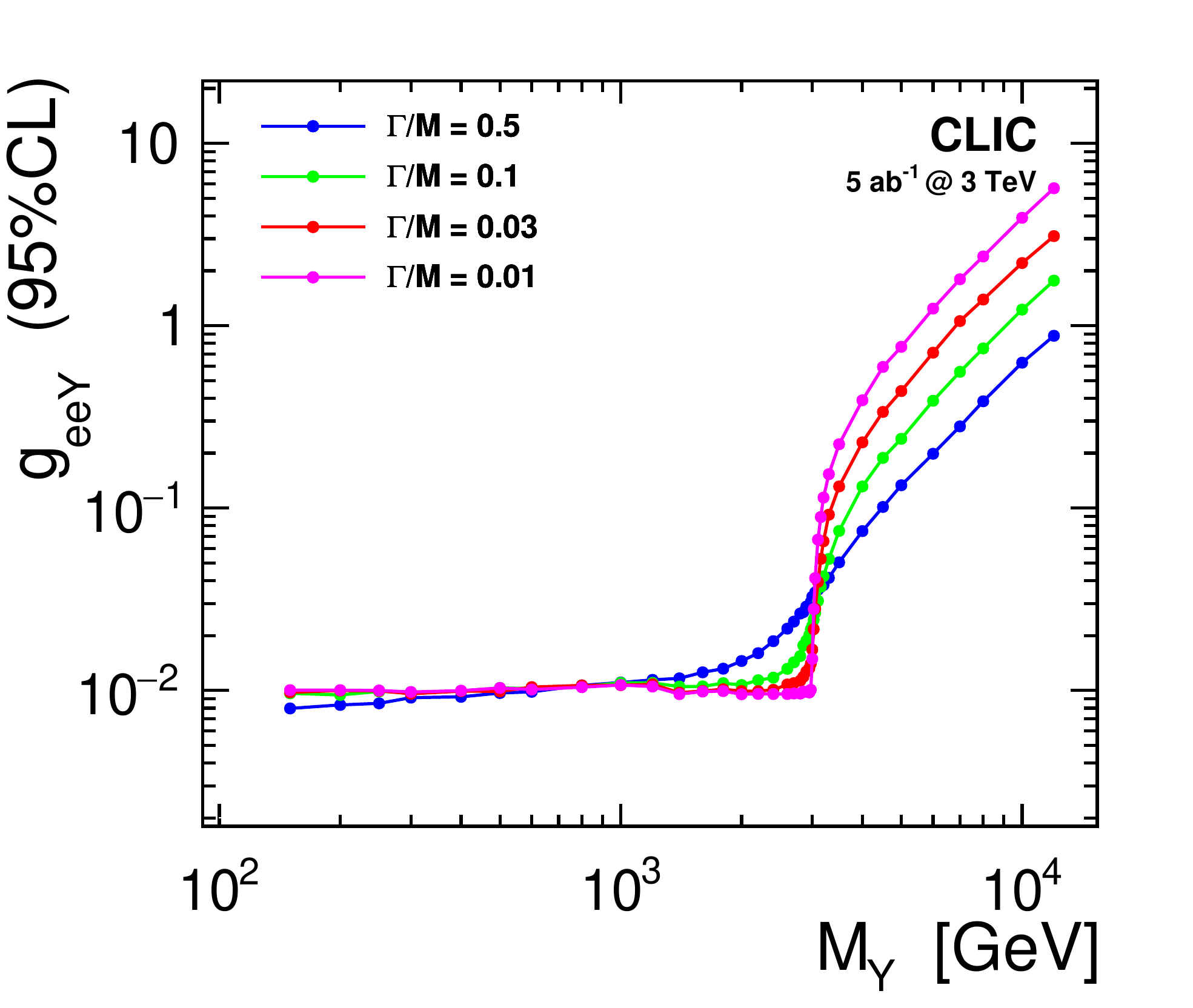}
  \caption{Limits on the vector mediator coupling to electrons for the
    ILC running at 500\,GeV (left) and CLIC running at 3\,TeV (right)
    and different fractional mediator widths, as indicated in the plot.
    Combined limits corresponding to the assumed running
    scenarios are presented with systematic uncertainties taken into
    account.
  } 
  \label{fig:coup_wid}
\end{figure}

As already discussed in  sec.~\ref{sec:strategy}, we assume that the
mediator coupling to SM particles is small and its decays to SM particles
can be neglected.
The total mediator width is dominated by the invisible width resulting
from its coupling to DM particle, $g_{\chi\chi Y}$. 
This coupling is thus fixed when considering a given
mediator scenario by specifying mediator mass and width.
This allows us to translate the extracted cross section limits to the
limits on the mediator coupling to electrons, $g_{eeY}$, 
the only free model parameter after fixing the DM type, coupling structure, masses
and mediator width. 
Limits on the mediator coupling to electrons, expected from combined
analysis of ILC and CLIC data,
are shown in Fig.~\ref{fig:coup_wid}
for different mediator widths.
For scenarios with light mediator exchange, SM
coupling limits hardly depend on the mediator width. For the Vector
mediator scenario, limits of $(1-4) \cdot 10^{-3}$ are expected at
500\,GeV ILC, improving with mediator mass (except for the largest
mediator width), while limit of about $10^{-2}$ is expected at 3\,TeV
CLIC, almost independent on the mediator mass (up to
kinematic limit) and width.
For the width of $\Gamma/M=0.03$ these coupling limits correspond to the
upper limits on the mediator branching ratio to electrons of about
$10^{-5}$ at 500\,GeV ILC and $10^{-4}$ at 3\,TeV CLIC.
This corresponds to single expected events and indicates that the
analysis of mono-photon spectra gives higher sensitivity to processes
with light mediator exchange than their direct searches in SM decay channels.

\begin{figure}[tbp]
 \includegraphics[width=0.49\textwidth]{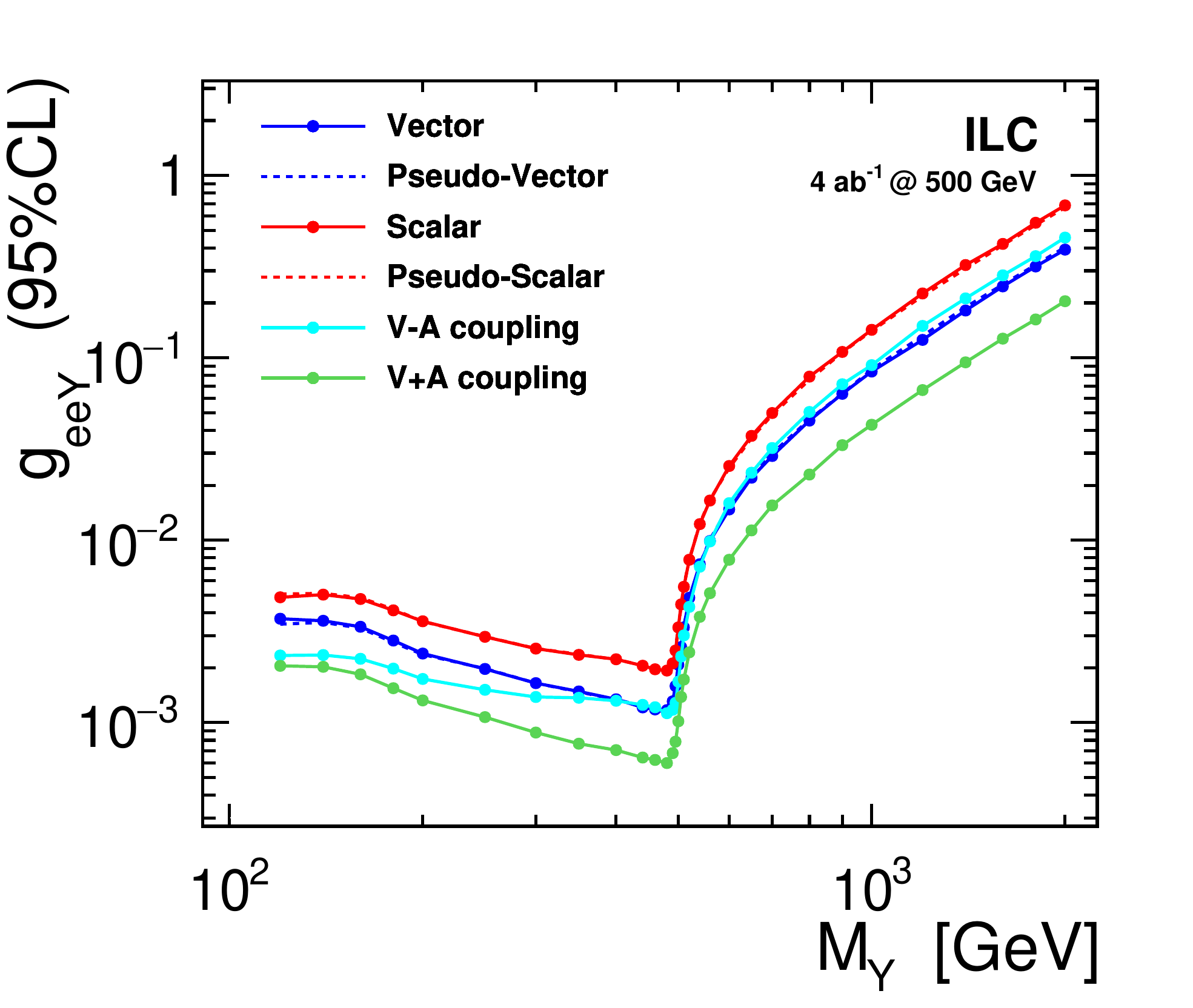}
 \includegraphics[width=0.49\textwidth]{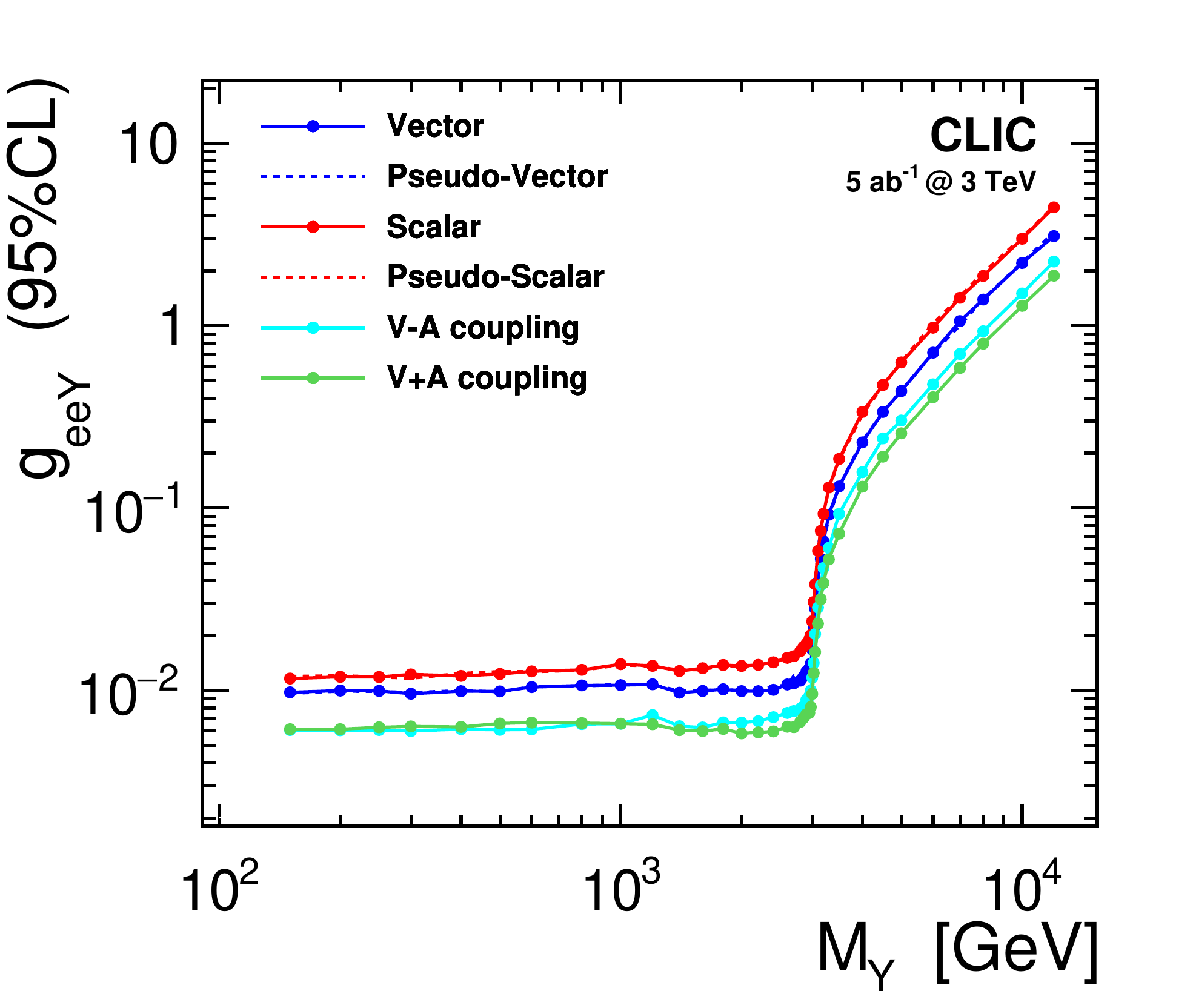}
  \caption{Limits on mediator coupling to electrons for the ILC
    running at 500\,GeV (left) and CLIC running at 3\,TeV (right), for
    relative mediator width, $\Gamma/M = 0.03$, and different mediator
    coupling scenarios, as indicated in the plot. Combined limits
corresponding to the assumed running scenarios are presented with
systematic uncertainties taken into account.  
  } 
  \label{fig:coup_model}
\end{figure}
As shown in Fig.~\ref{fig:coup_model},  coupling limits expected from
the combined analysis of all data vary by up to a factor of 3 for ILC
and up to a factor of 2 for CLIC, depending on the assumed coupling
structure. 
The strongest limits are obtained for the V+A mediator coupling
structure.
For this scenario the signal cross section is largest for polarisation
combinations corresponding to the lowest background levels.
The weakest limits are expected for Scalar and Pseudo-scalar coupling
scenarios.

\begin{figure}[tbp]
 \includegraphics[width=0.49\textwidth]{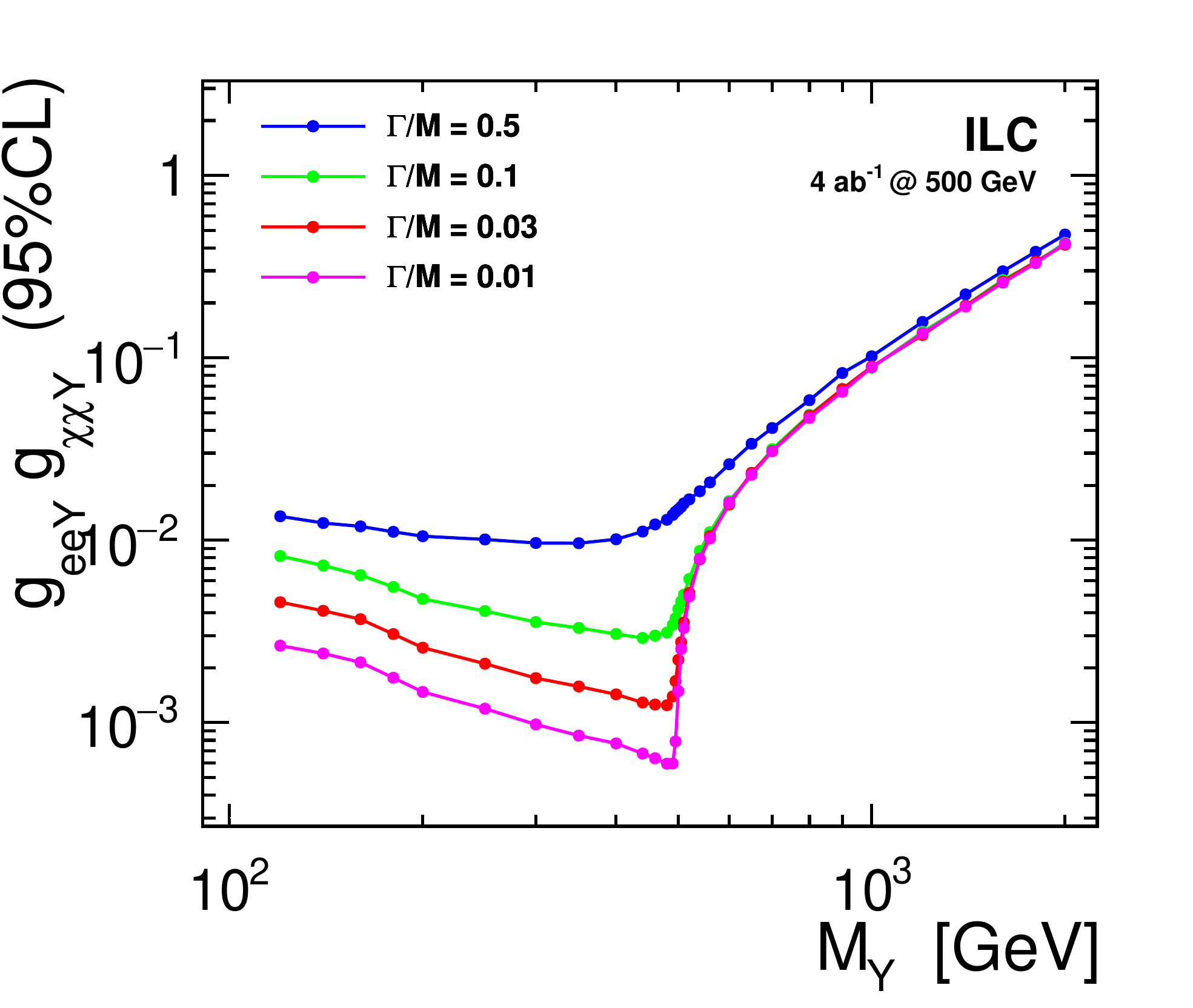}
 \includegraphics[width=0.49\textwidth]{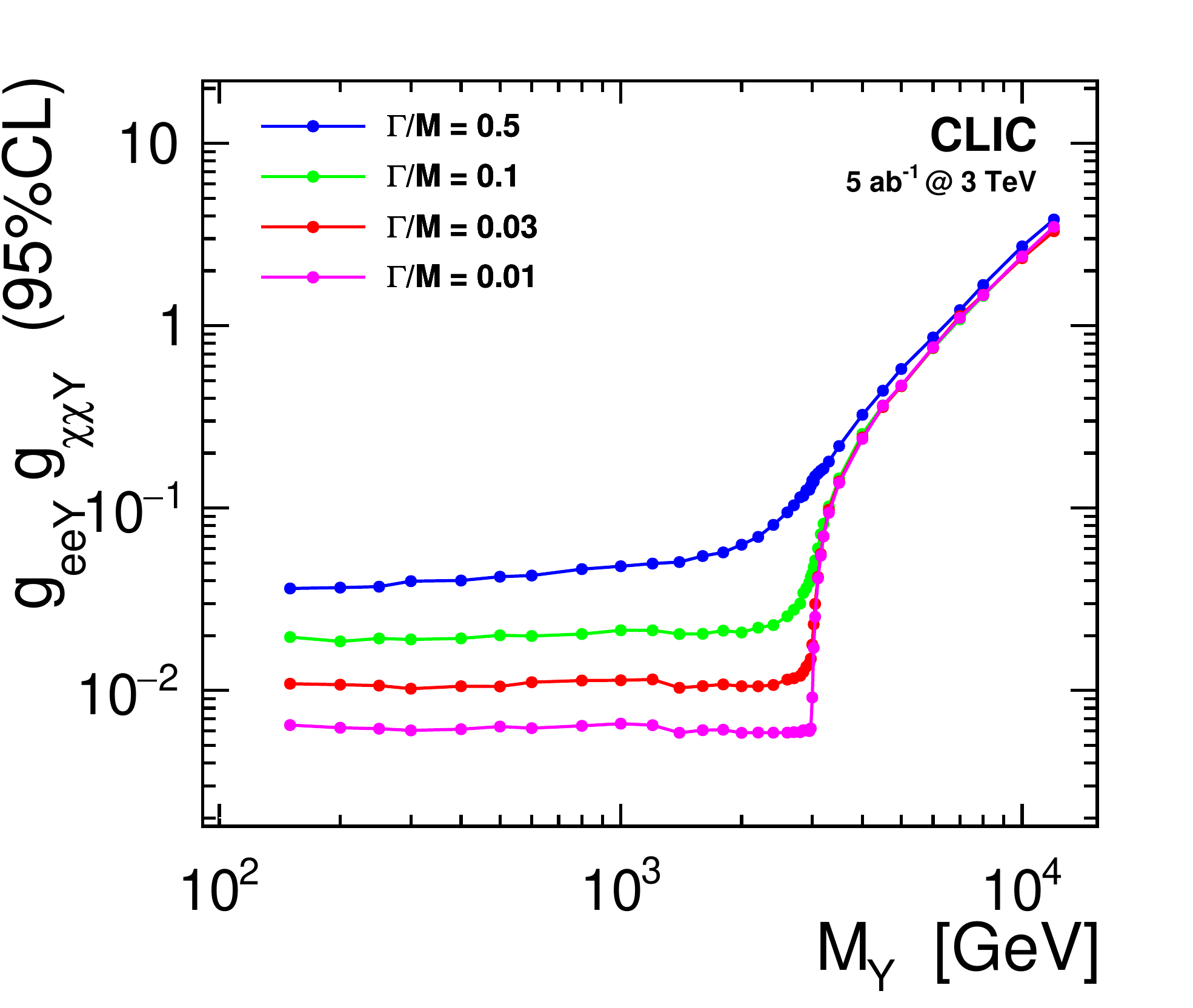}
  \caption{Limits on the product of the vector mediator couplings to
    electrons and to DM particles for the ILC running at 500\,GeV
    (left) and CLIC running at 3\,TeV (right) and different fractional
    mediator widths, as indicated in the plot. Combined limits
corresponding to the assumed running scenarios are presented with
systematic uncertainties taken into account. Pair-production
  of Dirac fermion DM is assumed. 
  } 
  \label{fig:coup2_model}
\end{figure}

For heavy mediator exchange, dark matter
pair-production cross section should no longer depend on the mediator
width.
However, results presented in Figs.~\ref{fig:coup_wid} show strong
width dependence in this mass range.
This is due to the fact that mediator coupling to DM particles,
$g_{\chi\chi Y}$, starts to be relevant for processes with exchange of
heavy (virtual) mediator and, as already discussed above, this
coupling is directly related to the mediator width.
Width dependence is removed when limits on the product of both couplings
are considered, $g_{eeY} \, g_{\chi\chi Y}$, as shown in
Fig.~\ref{fig:coup2_model}. 
In the limit of high mediator masses, $M_Y \gg \sqrt{s}$, limits on the
coupling product increase with square of the mediator mass.
This corresponds to the fixed limit on the effective mass scale of new
interactions, $\Lambda$, which would be obtained in the EFT approach:
\begin{eqnarray}
\Lambda^2 & = &  \frac{M^2_Y}{|g_{eeY} \, g_{\chi\chi Y}|} \, . \nonumber
\end{eqnarray}
For ILC running at 500\,GeV, expected EFT mass scale limits, resulting
from the combined analysis of all collected data, range from about
2.6\,TeV for Scalar mediator
and 3.1\,TeV for Vector mediator, in very
good agreement with full simulation results of \cite{Habermehl:2020njb},
to about 5.1\,TeV for V+A mediator
couplings.
For CLIC running at 3\,TeV, limits range from 6.1\,TeV and
  6.6\,TeV to 10.1\,TeV, respectively.


\section{Conclusions}

We propose a novel approach to estimate sensitivity of future $e^+
e^-$ colliders to  pair-production of DM particles via light mediator
exchange. 
The experimental sensitivity is defined in terms of the DM production
cross section as a function of both the mediator mass and mediator
width.
This approach is more general and more model-independent than the
approaches presented so far, assuming given mediator coupling values
to SM and DM particles.
We present expected limits for pair-production of light DM particles
at 500\,GeV ILC and 3\,TeV CLIC, based on analysis of the 2-D kinematic
distributions of mono-photon events. 
Expected limits on the radiative DM production cross section are at
the level of 1\,fb and weakly depend on the assumed mediator mass and
width.
Limits on the total DM pair-production cross section are of the order
of 10\,fb except for the resonant production region, $M_Y \sim
\sqrt{s}$, where hard photon radiation is significantly suppressed.
Extracted limits on the light mediator coupling to SM particles are in
the $10^{-3} - 10^{-2}$ range up to the kinematic limit,
$M_Y \le \sqrt{s}$.
For heavy mediator exchange, extracted cross section limits correspond
to the mass scale limits up to the order of 10\,TeV.

\section*{Acknowledgements}

We thank members of the CLIC detector and physics (CLICdp) collaboration
and the International Large Detector (ILD) concept group for the ILC
for fruitful discussions, valuable comments and suggestions. 
This contribution was supported by the National Science Centre, Poland,
the OPUS project under contract UMO-2017/25/B/ST2/00496 (2018-2021) and
the HARMONIA project under contract UMO-2015/\allowbreak18/M/\allowbreak{}ST2/\allowbreak00518 (2016-2021),
and by the German Research Foundation (DFG) under grant number STO 876/4-1
and STO 876/2-2.


\printbibliography

\end{document}